\documentclass[10pt,british,5p,times,fleqn,final]{elsarticle}
\biboptions{sort&compress}
\usepackage{fourier}

\usepackage[T1]{fontenc}
\usepackage{geometry}
\usepackage{color}
\usepackage{array}
\usepackage{textcomp}
\usepackage{amstext}
\usepackage{amssymb}
\usepackage{graphicx}
\usepackage{setspace}

\makeatletter

\providecommand{\tabularnewline}{\\}

\pdfmapfile{+txfonts.map} 
\usepackage{amsmath}
\usepackage{cases}

\journal{Materials \& Design}
\@ifundefined{showcaptionsetup}{}{%
 \PassOptionsToPackage{caption=false}{subfig}}
\usepackage{subfig}
\makeatother

\usepackage{babel}
\begin{document}
\global\long\def\crm{\mathrm{c}}
\global\long\def\im{\mathrm{i}}
\global\long\def\brm{\mathrm{b}}
\global\long\def\srm{\mathrm{s}}
\global\long\def\mrm{\mathrm{m}}
\global\long\def\erm{\mathrm{e}}
\global\long\def\irm{\mathrm{i}}
\global\long\def\Brm{\mathrm{B}}

\global\long\def\Frm{\mathrm{F}}
\global\long\def\visc{\mathrm{visc}}
\global\long\def\suprm{\mathrm{sup}}
\global\long\def\subrm{\mathrm{sub}}
\global\long\def\ths{\mathrm{th}}
\global\long\def\hyst{\mathrm{hyst}}
\global\long\def\eq{\mathrm{eq}}
\global\long\def\e{\mathrm{e}}
\global\long\def\Nset{\mathbb{N}}

\title{~\\
~\\
~\\
~\\
~\\
~\\
~\\
Optimisation of hybrid high-modulus/high-strength carbon fiber reinforced
plastic composite drive}

\author[crea,lma]{O.~Montagnier\corref{cor1}}

\ead{olivier.montagnier@inet.air.defense.gouv.fr}

\author[lma]{C.~Hochard}

\cortext[cor1]{Corresponding author. Tel. : (+33)4.90.17.80.93 ; fax : (+33)4.90.17.81.89.}

\address[crea]{Centre de recherche de l'Armée de l'air (CReA), Laboratoire de Dynamique
du Vol, BA 701, 13361 Salon Air, France}

\address[lma]{Laboratoire de Mécanique et d'Acoustique (LMA), 31 chemin Joseph
Aiguier, 13402 Marseille Cedex 20, France}
\begin{abstract}
This study deals with the optimisation of hybrid composite drive shafts
operating at subcritical or supercritical speeds, using a genetic
algorithm. A formulation for the flexural vibrations of a composite
drive shaft mounted on viscoelastic supports including shear effects
is developed. In particular, an analytic stability criterion is developed
to ensure the integrity of the system in the supercritical regime.
Then it is shown that the torsional strength can be computed with
the maximum stress criterion. A shell method is developed for computing
drive shaft torsional buckling. The optimisation of a helicopter tail
rotor driveline is then performed. In particular, original hybrid
shafts consisting of high-modulus and high-strength carbon fibre reinforced
epoxy plies were studied. The solutions obtained using the method
presented here made it possible to greatly decrease the number of
shafts and the weight of the driveline under subcritical conditions,
and even more under supercritical conditions. This study yielded some
general rules for designing an optimum composite shaft without any
need for optimisation algorithms.\end{abstract}
\maketitle
{\small }
\begin{figure*}[!t]
\begin{singlespace}
\centering{}{\small }%
\begin{minipage}[t]{0.49\textwidth}%
{\small{} }%
\begin{tabular}{>{\raggedright}p{1.7cm}p{5.6cm}}
\multicolumn{2}{l}{\textbf{\small Nomenclature}}\tabularnewline
{\small{} $\mathbf{A}$, $A_{ij}$ } & {\small in-plane stiffness matrix of the laminate and elements of
the matrix ($i,j=1,2,6$)}\tabularnewline
{\small{} $\mathbf{a}$, $a_{ij}$ } & {\small in-plane compliance matrix of the laminate and elements of
the matrix }\tabularnewline
{\small $\mathbf{B}$, $B_{ij}$ } & {\small coupling stiffness matrix of the laminate and elements of
the matrix}\tabularnewline
{\small{} $c$ } & {\small viscous damping }\tabularnewline
{\small $\mathbf{D}$, $D_{ij}$ } & {\small flexural stiffness matrix of the laminate and elements of
the matrix}\tabularnewline
{\small{} $E$} & {\small longitudinal Young's modulus of the shaft}\tabularnewline
{\small $E_{11}$, $E_{22}$, $E_{12}$, $E_{44}$, $E_{55}$, $E_{66}$} & {\small longitudinal and transverse Young's modulus, Poisson's ratio,
out-of-plane (transverse / normal, longitudinal / normal) and in-plane
shear modulus of the ply}\tabularnewline
{\small $f$} & {\small fitness function}\tabularnewline
{\small $G$} & {\small transverse shear modulus of the shaft}\tabularnewline
{\small $g$} & {\small constraint function}\tabularnewline
{\small $h$} & {\small number of half-wave along the circumference}\tabularnewline
{\small{} $I_{y}$, $I_{z}$ } & {\small transverse area moments of inertia of the shaft}\tabularnewline
{\small{} $J$ } & {\small polar mass moment of inertia}\tabularnewline
{\small{} $\mathbf{K}$ } & {\small buckling stiffness matrix}\tabularnewline
{\small{} $K$ } & {\small reserve factors}\tabularnewline
{\small{} $k$ } & {\small stiffness }\tabularnewline
{\small{} $l$ } & {\small unsupported shaft section length}\tabularnewline
{\small{} $m$ } & {\small mass }\tabularnewline
{\small{} $n$ } & {\small number of plies}\tabularnewline
{\small{} $P$ } & {\small power}\tabularnewline
{\small $p$} & {\small number of half-wave along the axis}\tabularnewline
{\small{} $q$ } & {\small number of possible orientations in the staking sequence}\tabularnewline
{\small{} $r$ } & {\small shaft radius}\tabularnewline
{\small{} $S$ } & {\small cross-section area of the shaft }\tabularnewline
{\small{} $t$ } & {\small thickness or time}\tabularnewline
{\small{} $T$ } & {\small axial torque}\tabularnewline
{\small{} $u$, $v$, $w$ } & {\small displacements (complex or real)}\tabularnewline
{\small $X$, $X'$, $Y$, $Y'$, $s$} & {\small tensile and compressive longitudinal strength, tensile and
compressive transverse strength, in-plane shear strength }\tabularnewline
\end{tabular}%
\end{minipage}{\small }%
\begin{minipage}[t]{0.49\textwidth}%
{\small{} }%
\begin{tabular}{>{\raggedright}p{1.7cm}p{5.6cm}}
{\small{} $\mathbf{x}$, $\mathbf{y}$, $\mathbf{z}$} & {\small coordinates}\tabularnewline
{\small{} $\alpha$ } & {\small orientation of the ply }\tabularnewline
{\small{} $\eta$ } & {\small loss factor }\tabularnewline
{\small{} $\kappa$ } & {\small shear coefficient }\tabularnewline
{\small $\nu$ } & {\small Poisson's ratio of the shaft}\tabularnewline
{\small $\upsilon$ } & {\small distribution of the torsional modes}\tabularnewline
{\small $\rho$ } & {\small mass density of the shaft }\tabularnewline
{\small $\varphi$ } & {\small angular position}\tabularnewline
{\small $\theta$} & {\small out-of plane cross-section rotation (complex or real)}\tabularnewline
{\small $\sigma_{11}$, $\sigma_{22}$, $\sigma_{12}$} & {\small in-plane stress of the ply (longitudinal, transverse and shear)}\tabularnewline
{\small $\varpi$ } & {\small natural frequency of the torsional modes}\tabularnewline
{\small{} $\omega$ } & {\small natural frequency of the flexural modes }\tabularnewline
{\small{} $\Omega$ } & {\small spin speed}\tabularnewline
 & \tabularnewline
{\small{} }\textit{\small Subscript} & \tabularnewline
{\small{} $\Brm-$, $\Brm+$, } & {\small lower and higher backward whirl speeds }\tabularnewline
{\small{} $\brm$ } & {\small bearing}\tabularnewline
{\small{} $\mathrm{buck}$ } & {\small buckling}\tabularnewline
{\small{} $\crm$ } & {\small critical}\tabularnewline
{\small{} $\mathrm{dv}$ } & {\small driveline}\tabularnewline
{\small{} $\erm$ } & {\small external}\tabularnewline
{\small{} $\eq$ } & {\small equivalent}\tabularnewline
{\small{} $\Frm-$, $\Frm+$ } & {\small lower and higher forward whirl speeds }\tabularnewline
{\small{} $\mathrm{f}$} & {\small flexural modes}\tabularnewline
{\small{} $\mathrm{G}$ } & {\small gear}\tabularnewline
{\small{} $\irm$ } & {\small internal}\tabularnewline
{\small{} $\mathrm{inf}$ } & {\small inferior}\tabularnewline
{\small{} $\mathrm{m}$ } & {\small medium }\tabularnewline
{\small{} $\mathrm{min}$ } & {\small minimum }\tabularnewline
{\small{} $\mathrm{mat}$ } & {\small material of a ply}\tabularnewline
{\small{} $\mathrm{nom}$ } & {\small nominal }\tabularnewline
{\small{} $n$ } & {\small number of sine modes}\tabularnewline
{\small{} $\mathrm{ply}$ } & {\small laminate ply}\tabularnewline
{\small{} $\srm$ } & {\small shaft}\tabularnewline
{\small{} $\suprm$ } & {\small superior}\tabularnewline
{\small{} $\mathrm{T}$ } & {\small tail rotor}\tabularnewline
{\small{} $\mathrm{t}$ } & {\small torsional modes}\tabularnewline
{\small{} $\ths$ } & {\small threshold speed}\tabularnewline
{\small{} $\mathrm{str}$} & {\small strength}\tabularnewline
\end{tabular}%
\end{minipage}\end{singlespace}
\end{figure*}
{\small \par}

\section{Introduction }

{\small Since the 1970s, composite materials have been regarded as
potential candidates for manufacturing drive shafts because of their
high specific stiffness and strength \citep{Zinberg70}. Previous
studies on this topic have dealt mainly with composite shaft design
in the subcritical case, that is when the first critical speed is
never exceeded. However, when a long driveline is required (in the
case of helicopters, tilt-rotors, tailless aircraft with twin turboprops,
etc.), an additional means of increasing the drive shaft length consists
in operating above the first critical speed, in the so-called supercritical
regime. The main advantage of long shafts is that they reduce the
number of bearing supports required, and thus greatly decrease the
maintenance costs and the weight of the driveline. The design process
is more complex, however, because the shaft has to cross a critical
speed, and dynamic instabilities due to rotating damping can occur
in this regime. Aeronautic applications lend themselves well to operating
in the supercritical regime because the driveline always rotates at
the nominal speed during flight, since they undergo acceleration and
deceleration processes on the ground. The aim of this paper is to
optimise a supercritical drive shaft in this practical case. }{\small \par}

{\small Many different numerical methods have been used to design
optimised composite drive shafts in order to reduce their weight,
for example. Traditional methods based on the gradients of continuous
functions have been used for this purpose by several authors \citep{Bauchau83,Lim86,Darlow95}.
These methods are unsuitable in the case of composite laminates, however,
because many of the variables which have to be optimised are discrete
variables (such as the number of plies and the ply angle in prepreg
lay-up processes). It is therefore necessary to assume these variables
to be continuous to be able to compute the gradients required. The
optimisation techniques available for solving problems involving discrete
variables are known as metaheuristic methods. For example, \citet{Gubran02a}
have used simulated annealing techniques based on a neighbourhood
approach. A review of the literature shows that genetic algorithms
(GA) \citep{Goldberg89,Awad12} are well adapted to designing laminate
structures. GA were recently used to optimise a flexible matrix composite
drive shaft in \citep{Roos11}. Here it is proposed to use a GA with
penalisation methods to account for the constraint functions. In addition,
in order to reduce the CPU time, all the design aspects are handled
without requiring the use of finite element methods.}{\small \par}

{\small In drive shaft applications, the choice of composite material
is of great importance. Several authors have recommended the use of
hybrid composites in the production of drive shafts. \citet{Xu91},
\citet{Gubran05b}, and \citet{Badie11} recently studied the advantages
of a mixture of glass and carbon fibres in a modified epoxy matrix.
\citet{Lee04}, \citet{Gubran05b}, \citet{Mutasher09} and \citet{Abu10}
recently have studied the design and manufacture of hybrid metallic/composite
drive shafts. Here it is proposed to study the use of a combination
of high-modulus (HM) and high-strength (HS) carbon fibre reinforced
epoxy plies, in order to benefit from the main advantages of each
type of fibre. The main design considerations in the case of composite
laminate tubes are the axial stiffness and the torsional resistance.
In this particular case, the plies providing stiffness and strength
can be considered practically separately \citep{HochardCST05}. HM
car\-bon/e\-po\-xy, which have poor strength properties, especially
when exposed to compression loads \citep{MontagnierJCM05}, can serve
to maximise the axial stiffness. HS car\-bon/e\-po\-xy can be used
to maximise the resistance to torsion loads. Note that the hybridization
can be simply obtained by using the same resin with both fibres. Otherwise,
it would be necessary to verify the compatibility of the two resins
(in terms of their curing cycle behaviour, adhesiveness, etc.). For
the sake of convenience, this point is not taken into account here. }{\small \par}

{\small The first part of this paper presents various design aspects.
In particular, the dynamic of a drive shaft is studied at supercritical
speeds. The failure strength analysis focuses on the choice of the
stress criterion. A composite shell model is then developed for the
torsional buckling. In the second part, the GA is presented. The last
part presents a comparative study between HM and hybrid solutions
on a helicopter tail rotor driveline previously presented in the literature.}{\small \par}

\section{Design aspects}

\subsection{Flexural vibration analysis}

\begin{figure*}[t]
\begin{centering}
{\footnotesize \includegraphics[width=13cm]{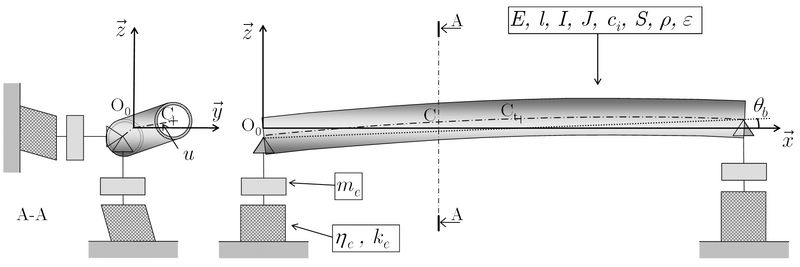}}
\par\end{centering}{\footnotesize \par}

{\footnotesize \caption{{\footnotesize A simply supported axisymmetric tubular composite shaft
with rolling-element bearings mounted on viscoelastic supports \label{fig:arbre_supports}}}
}
\end{figure*}

{\small When designing supercritical shafts, the external damping
has to be maximised in order to reduce the flexural imbalance responses
and increase the stability in the supercritical regime. Rolling-element
bearings provide insufficient damping. Dissipative materials such
as elastomers have recently been used as bearing supports as a passive
means of enhancing the non-rotating damping \citep{Dutt96,MontagnierJSV07}.
A low cost configuration consisting of an axisymmetric composite shaft
simply supported on classical rolling-element bearings mounted on
viscoelastic supports was studied here (Fig.~\ref{fig:arbre_supports}). }{\small \par}

\begin{table*}[t]
\caption{{\footnotesize Material properties corresponding to a volume fraction
of 0.6 \label{tab:carbon_properties}}}

\begin{centering}
\begin{tabular}{lcccccccccccc}
\hline 
{\footnotesize Material} & {\footnotesize Abbr.} & {\footnotesize $\rho$} & {\footnotesize $E_{11}$} & {\footnotesize $E_{22}$} & {\footnotesize $E_{66}$} & {\footnotesize $E_{12}$} & {\footnotesize $X$} & {\footnotesize $X'$} & {\footnotesize $Y$} & {\footnotesize $Y'$} & {\footnotesize $s$} & {\footnotesize $t_{\mathrm{ply}}$ }\tabularnewline
\hline 
 &  & {\footnotesize kg\,m$^{-3}$} & {\footnotesize GPa} & {\footnotesize GPa} & {\footnotesize GPa} & {\footnotesize -} & {\footnotesize MPa} & {\footnotesize MPa} & {\footnotesize MPa} & {\footnotesize MPa} & {\footnotesize MPa} & {\footnotesize mm}\tabularnewline
\hline 
{\footnotesize Narmco 5505 \citep{Zinberg70}} & {\footnotesize BE}\textcolor{black}{\footnotesize $^{a}$} & {\footnotesize 1\,965} & {\footnotesize 211} & {\footnotesize 24.1} & {\footnotesize 6.89} & {\footnotesize 0.36} & {\footnotesize 1\,365} & {\footnotesize 1\,586} & {\footnotesize 45} & {\footnotesize 213} & {\footnotesize 62} & {\footnotesize 0.1321}\tabularnewline
{\footnotesize T300/5208 \citep{Lim86,Darlow95}} & {\footnotesize CE$_{\mathrm{L.}}$} & {\footnotesize 1\,680} & {\footnotesize 181} & {\footnotesize 10.3} & {\footnotesize 7.17} & {\footnotesize 0.28} & {\footnotesize 1\,500} & {\footnotesize 1\,500} & {\footnotesize 40} & {\footnotesize 246} & {\footnotesize 68} & \tabularnewline
{\footnotesize K63712/M10 \citep{MontagnierJCM05}} & {\footnotesize HM} & {\footnotesize 1\,700} & {\footnotesize 370} & {\footnotesize 5.4} & {\footnotesize 4.0} & {\footnotesize 0.3} & {\footnotesize 1\,500} & {\footnotesize 470} & {\footnotesize 35} & {\footnotesize 200} & {\footnotesize 75} & {\footnotesize 0.125}\tabularnewline
{\footnotesize T800/G947} & {\footnotesize HS} & {\footnotesize 1\,530} & {\footnotesize 162} & {\footnotesize 10} & {\footnotesize 5.0} & {\footnotesize 0.3} & {\footnotesize 2\,940} & {\footnotesize 1\,570} & {\footnotesize 60} & {\footnotesize 290} & {\footnotesize 100} & {\footnotesize 0.125}\tabularnewline
\hline 
\end{tabular}
\par\end{centering}

\begin{centering}
\textcolor{black}{\footnotesize $^{a}$~BE : boron/epoxy ; CE : carbon/epoxy}
\par\end{centering}{\footnotesize \par}

\end{table*}

{\small }{\small \par}

{\small Various approaches based on beam and shell theories have been
used to compute the critical speeds of composite shafts \citep{BertKim95,Gubran05,Sino08},
most of which were placed on infinitely rigid supports. The simplest
of these theories is called the Equivalent Modulus Beam Theory}\linebreak{}
{\small (EMBT) \citep{Zinberg70}. Based on this approach, it is proposed
to investigate a rotating beam with Timoshenko's assumptions \citep{Timoshenko37},
replacing the isotropic properties of the material by the homogenised
properties of the composite. These equations are also adapted to account
for the motion of the supports and the internal damping terms. Lastly,
the three complex governing equations and boundary conditions used
can be written in the following form:}{\small \par}

{\small 
\begin{gather}
\ddot{u}-\frac{I_{y}}{S}\left(1+\frac{E}{\kappa G}\right)\ddot{u_{s}}''+\im\Omega\frac{I_{x}}{S}\dot{u_{\srm}}''\nonumber \\
+\frac{EI_{y}}{\rho S}u_{\srm}''''+\frac{c_{i}}{\rho Sl}\left(\dot{u}_{s}-\im\Omega u_{s}\right)=0\ ,\label{eq:ray_tim_final}\\
\int\nolimits _{0}^{l}\rho S\ddot{u}\mathrm{d}x+2m_{\brm}\ddot{u}_{\brm}+2c_{\erm}\dot{u}_{\brm}+2k_{\erm}u_{\brm}=0\ ,\\
\int\nolimits _{0}^{l}\rho S\left(x-\frac{l}{2}\right)\ddot{u}\mathrm{d}x+2m_{\brm}\frac{l^{2}}{4}\ddot{\theta}_{\brm}+2c_{\erm}\frac{l^{2}}{4}\dot{\theta}_{\brm}+2k_{\erm}\frac{l^{2}}{4}\theta_{\brm}=0\ ,\\
u''_{\srm}(0,t)=\ u''_{\srm}(l,t)=0,\ u_{\srm}(0,t)=\ u_{\srm}(l,t)=0
\end{gather}
where $\_'=\partial\_/\partial x$, $\dot{\_}=\partial\_/\partial t$,
$\rho=m_{\srm}/Sl$, $u=u_{y}+\im u_{z}$ is the cross-sectional displacement
and $u_{\brm}=u_{\brm y}+\im u_{\brm z}$ is the deflection of the
shaft (see the list of nomenclature for the other parameters). Using
the method presented in \citep{MontagnierJSV07}, the above equations
yield the four critical speeds for the $n\mathrm{th}$ harmonic:
\begin{gather}
\omega_{\crm n\Frm\pm}=\frac{1}{\sqrt{2\Delta_{n-}}}\left[\omega_{\srm n}^{2}+\Lambda_{n-}\omega_{\brm n}^{2}\right.\nonumber \\
\left.\pm\sqrt{\omega_{\srm n}^{4}+2(\Lambda_{n-}-2\Delta_{n-})\omega_{\srm n}^{2}\omega_{\brm n}^{2}+\Lambda_{n-}^{2}\omega_{\brm n}^{4}}\right]^{\frac{1}{2}}\label{eq:vitessecritiqueF}
\end{gather}
\begin{gather}
\omega_{\crm n\Brm\pm}=-\frac{1}{\sqrt{2\Delta_{n+}}}\left[\omega_{\srm n}^{2}+\Lambda_{n+}\omega_{\brm n}^{2}\right.\nonumber \\
\left.\pm\sqrt{\omega_{\srm n}^{4}+2(\Lambda_{n+}-2\Delta_{n+})\omega_{\srm n}^{2}\omega_{\brm n}^{2}+\Lambda_{n+}^{2}\omega_{\brm n}^{4}}\right]^{\frac{1}{2}}\label{eq:vitessecritiqueB}
\end{gather}
where
\begin{gather}
\omega_{\srm n}^{2}=\frac{n^{4}\pi^{4}EI_{y}}{\rho Sl^{4}}=\frac{k_{\srm n}}{m_{\srm}},\;\omega_{\brm n}^{2}=\frac{k_{\erm}}{m_{\brm}+\frac{m_{\srm}}{2(2+(-1)^{n})}},\;\Gamma_{n}=\frac{n^{2}\pi^{2}I_{x}}{Sl^{2}},\nonumber \\
\Pi_{n}=1+\frac{n^{2}\pi^{2}I_{y}}{Sl^{2}}\left(1+\frac{E}{\kappa G}\right),\;\Phi_{n}=\frac{m_{\srm}}{m_{\brm}+\frac{m_{\srm}}{2(2+(-1)^{n})}},\nonumber \\
\Psi_{n}=\Pi_{n}-\frac{4}{n^{2}\pi^{2}}\Phi_{n},\;\Delta_{n\pm}=\Psi_{n}\pm\Gamma_{n},\;\Lambda_{n\pm}=\Pi_{n}\pm\Gamma_{n}\label{eq:parametres}
\end{gather}
}{\small \par}

{\small In addition, in the case of a composite shaft consisting of
a symmetric laminate, the homogenised properties can be computed with
the following equations: $E=1/a_{11}t_{\srm}$, $G=1/a_{66}t_{\srm}$
and $\nu=-a_{12}/a_{11}$ where $\mathbf{a}=\mathbf{A}^{-1}$ \citep{Tsai80}.
It is also assumed that $\kappa=2(1+\nu)/(4+3\nu)$.}{\small \par}

{\small In the field of rotordynamics, internal damping, which is
also referred to as rotating damping, is known to cause whirl instability
in the supercritical regime. In the literature, the internal damping
resulting from dissipation in the shaft material and dry friction
between the assembled components has been usually approached using
the viscous damping model. However, most materials, such as car\-bon/e\-po\-xy
materials in particular, show vibratory damping, which resembles hysteretic
damping much more than viscous damping \citep{Adams87,MontagnierPhd}.
Using the classical equivalence between viscous and hysteretic damping
\citep{MontagnierJSV07}, the analytical instability criterion suitable
for shaft optimisation purposes, can be written in the following form:}{\small \par}

{\small 
\begin{gather}
\pm\left(\eta_{\erm}k_{\erm}\Phi_{n}(\Pi_{n}\omega_{n\mathrm{F\pm}0}^{2}-\omega_{\srm n}^{2})-\eta_{\irm}k_{\srm n}(\omega_{n\mathrm{F\pm}0}^{2}-\omega_{\brm n}^{2})\right)\nonumber \\
\begin{cases}
<0 & \Longrightarrow\ \omega_{\ths\, n\Frm\pm}=\omega_{n\Frm\pm0}\\
\geqslant0 & \textrm{\ensuremath{\Longrightarrow}\ \textrm{stable}}
\end{cases}\label{eq:lim.hystplus}
\end{gather}
where
\begin{gather}
\omega_{n\mathrm{F\pm}0}=\frac{1}{\sqrt{2\Psi_{n}}}\left[\omega_{\srm n}^{2}+\Pi_{n}\omega_{\brm n}^{2}\right.\nonumber \\
\left.\pm\sqrt{\omega_{\srm n}^{4}+2(\Pi_{n}-2\Psi_{n})\omega_{\srm n}^{2}\omega_{\brm n}^{2}+\Pi_{n}^{2}\omega_{\brm n}^{4}}\right]^{\frac{1}{2}}\label{eq:pulsation41}
\end{gather}
}{\small \par}

{\small and where the equivalent longitudinal loss factor denoted
$\eta_{\irm}$ is computed with Adams, Bacon and Ni's theory \citep{Adams73,Ni84}
using complex properties of the ply ($\eta_{11}=$0.11\,\%, $\eta_{22}=$0.70\,\%
and $\eta_{66}=$1.10\,\%). It then suffices to compute the lowest
threshold speed in order to determine the spin speed limit of the
shaft. }{\small \par}

\begin{figure}
\begin{centering}
\includegraphics[width=9cm]{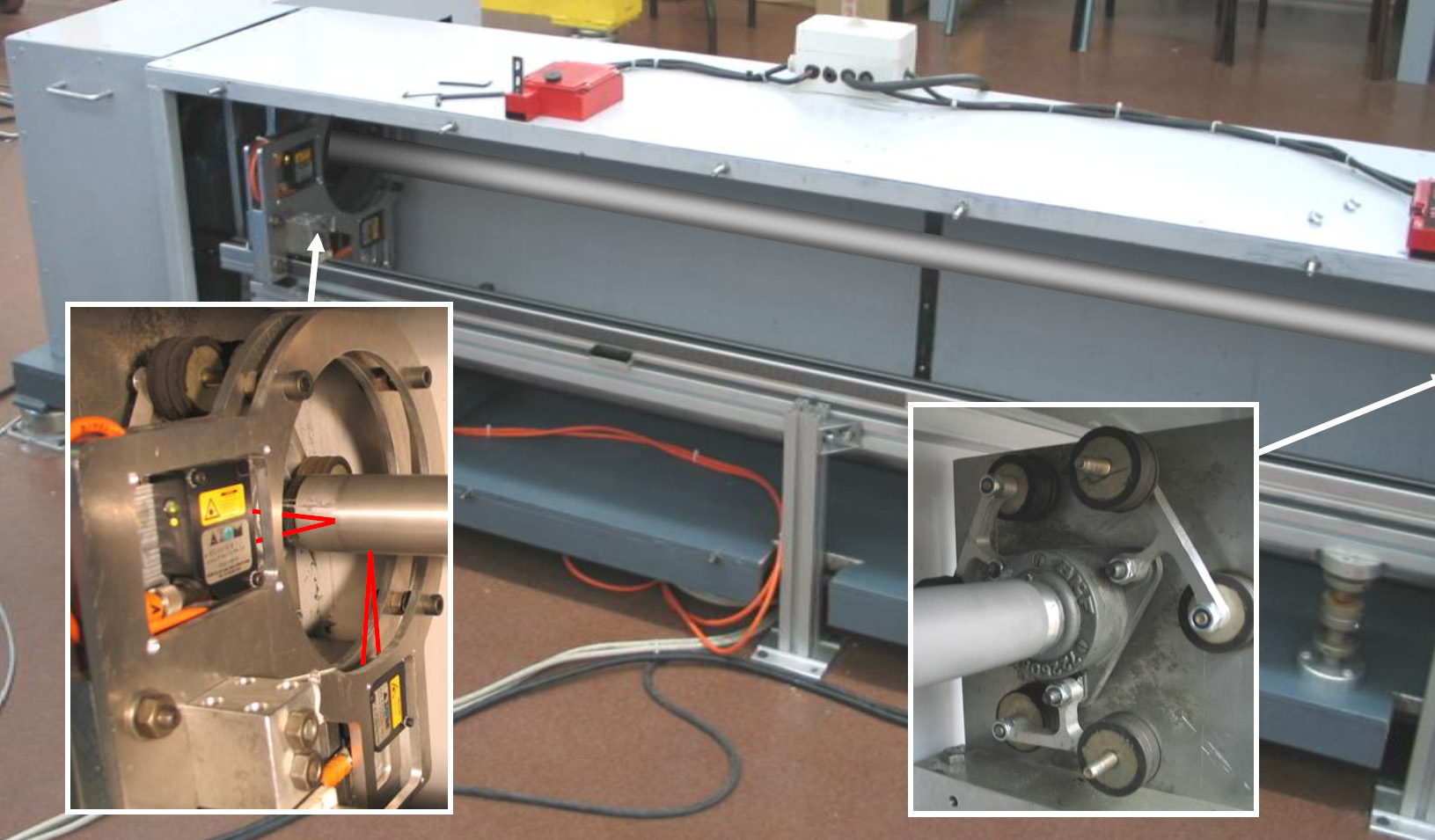}
\par\end{centering}

\caption{{\footnotesize The dynamic shaft test rig: the zoom image (left) corresponds
to the non-contact laser-optical displacement sensors (with the scheme
of the laser beams in red), the detail (right) corresponds to the
bearing mounted on the six viscoelastic supports (the left plate was
removed to take the snapshot).\label{fig:dynamic-test-rig} }}
\end{figure}

{\small A dynamic shaft test rig corresponding to the case of Fig.~\ref{fig:arbre_supports}
has been developed to validate the theoretical results (Fig.~\ref{fig:dynamic-test-rig}).
Basically, the test rig consists in a shaft that is powered by an
electric motor via a belt and pulley system and is capable of a maximum
test velocity of 12\,000\,rpm. Two non-contact laser-optical displacement
sensors are able to measure the cross-section displacement in real-time.
 The detailed characteristics of the test rig are given in \citep{MontagnierPhd,MontagnierICRD06}.
In the case of a long aluminium shaft ($E=69$\,GPa, $\rho=2700$\,kg.m$^{-1}$,
$l=1.80$\,m, $r_{\mrm}=23.99$\,mm, $t=2.02$\,mm) supported on
viscoelastic supports ($m_{\brm}=2.817$\,kg, $k_{\erm}=5.64\times10^{5}$\,N.m$^{-1}$),
the critical speeds $\omega_{\crm1\Frm-}$ and $\omega_{\crm1\Frm+}$
were measured to be $251$\,rad.s$^{-1}$ and $446$\,rad.s$^{-1}$,
respectively.  The results obtained with the above model are $250$\,rad.s$^{-1}$
and $460$\,rad.s$^{-1}$ which is very close to the experiment.
}{\small \par}

\begin{figure}
\begin{centering}
\includegraphics[width=9cm]{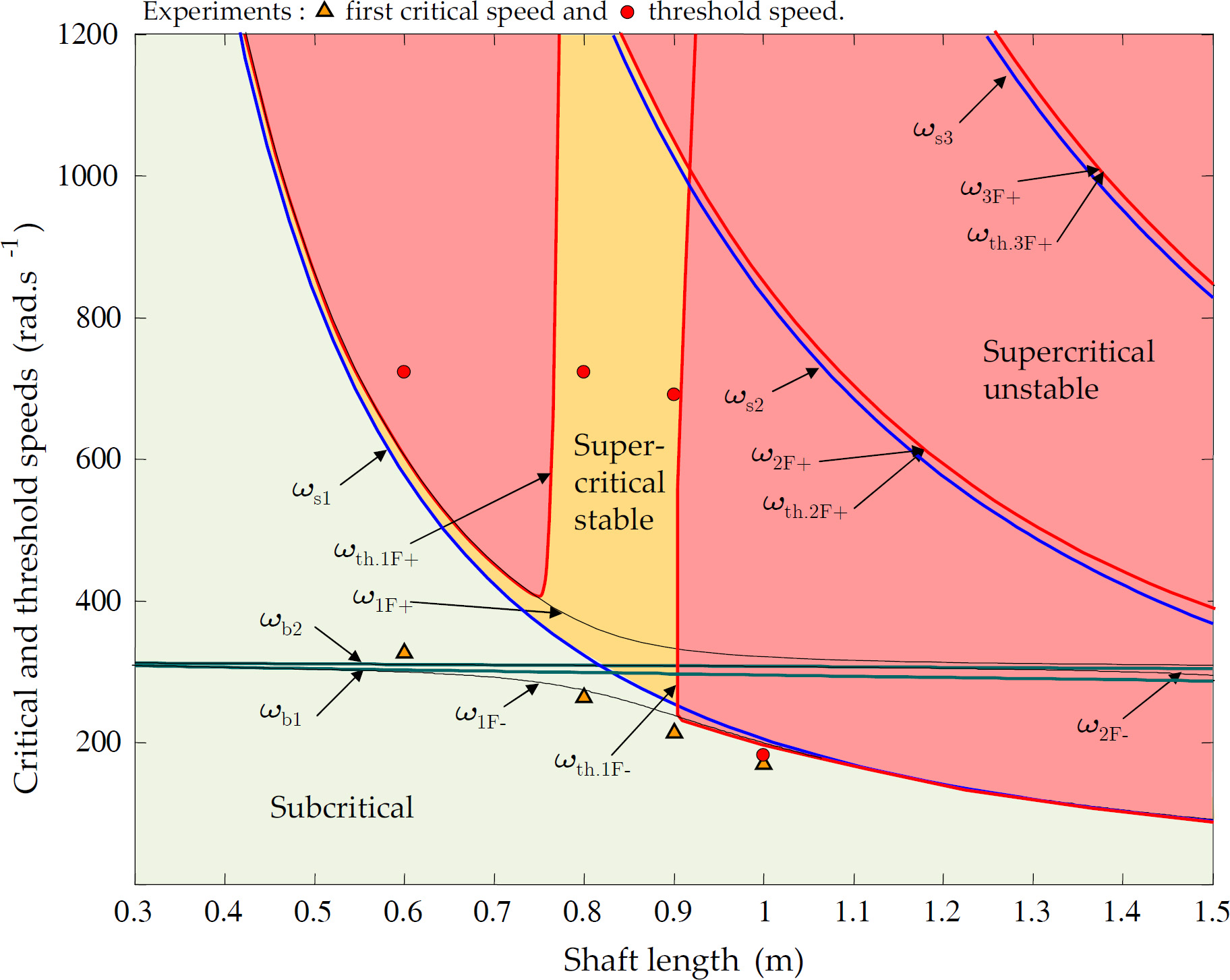}
\par\end{centering}

\caption{{\footnotesize The theoretical uncoupled natural frequencies ($\omega_{\srm n}$,
$\omega_{\brm n}$), forward critical speeds ($\omega_{n\Frm\pm}$)
and threshold speeds ($\omega_{\ths.n\Frm\pm}$) vs. experimental
data in the case of tubes in PVC material of various length.\label{fig:instability}}}

\end{figure}

{\small The experimental investigation of the instabilities can initiate
catastrophic risks for the dynamic test rig. For this reason, it was
proposed to study the instabilities using PVC material. Another advantage
of the PVC material is its low stiffness and high damping ($E=2.2$\,GPa,
$\rho=1350$\,kg.m$^{-1}$ and $\eta_{\irm}=0.025$\,\%) which decrease
the critical and threshold speeds. Several shafts with four different
lengths were tested in the supercritical domain ($r_{\mrm}=23.25$\,mm,
$t=2.5$\,mm with $m_{\brm}=2.608$\,kg, $k_{\erm}=2.58\times10^{5}$\,N.m$^{-1}$
and $\eta_{\erm}=0.07$\,\%). A high level of acceleration was required
in order to run over the first critical speed. Only the shafts of
length 0.8\,m and 0.9\,m were found to be stable in the supercritical
regime. The theoretical model is compared to experimental results
in the Fig.~\ref{fig:instability}. A relatively good correlation
is obtained considering these experiments are difficult to achieve.
Note that the stability area is obtained at the intersection of the
rigid body mode and the first flexural mode ($l=0.85$\,m).}{\small \par}

\subsection{Torsional vibration analysis}

{\small Torsional vibrations are computed using classical methods
with the relations presented by \citet{Lim86}:
\begin{gather}
\varpi_{n}=\frac{\upsilon_{n}}{l}\sqrt{\frac{G}{\rho}}\label{eq:freq_torsion-1-1}
\end{gather}
with}{\small \par}
{\small 
\begin{gather*}
\upsilon_{1}\approx\sqrt{2}\sqrt{\frac{J_{\mathrm{G}}J_{\srm}+J_{\mathrm{T}}J_{\srm}+J_{\srm}^{2}}{J_{\mathrm{G}}J_{\srm}+J_{\mathrm{T}}J_{\srm}+2J_{\mathrm{G}}J_{\mathrm{T}}}},
\end{gather*}
\begin{gather*}
\upsilon_{n\neq1}\approx\frac{(n-1)\pi}{2}+\sqrt{\frac{(n-1)^{2}\pi^{2}}{4}+\frac{J_{\srm}}{J_{\mathrm{T}}}+\frac{J_{\srm}}{J_{\mathrm{G}}}}
\end{gather*}
where $J_{\mathrm{G}}$, $J_{\mathrm{T}}$ and $J_{\mathrm{s}}$ are
the mass moment of inertia of the main gearing, the tail rotor and
the shaft, respectively.}{\small \par}

\subsection{Failure strength analysis}

{\small Only the torsional resistance of the shaft is taken into account
here. The transmitted torque causes only in-plane shear, which can
be computed with classical laminate theory \citep{Tsai80}. Contrary
to what occurs with an unsymmetrical free-edge laminate plate, the
tubular structure blocks the coupling effects in the case of small
displacements. This can be modelled simply by assuming the classical
coupling matrix $\mathbf{B}$ to be null before performing the inversion
procedure required to compute the strain state. }{\small \par}

{\small A conservative approach often used in the case of helicopter
drive shafts consists in computing only the fracture of the first
ply. The Tsai-Wu criterion \citep{TsaiWu71} can be used in this case
to account for the differences between the tensile and compressive
strengths, which can be of great importance in the case of HM car\-bon/e\-po\-xy
(see Table\,\ref{tab:carbon_properties}): 
\begin{gather}
\frac{1}{XX'}\sigma_{11}^{2}+2\frac{F_{12}}{\sqrt{XX'YY'}}\sigma_{11}\sigma_{22}+\frac{1}{YY'}\sigma_{22}^{2}\nonumber \\
+\frac{1}{s^{2}}\sigma_{12}^{2}+(\frac{1}{X}-\frac{1}{X'})\sigma_{11}+(\frac{1}{Y}-\frac{1}{Y'})\sigma_{22}\leq1
\end{gather}
where $F_{12}$ is an interaction parameter which is taken to be equal
to 0.5. It should be noted that the Tsai-Wu criterion includes the
transverse fracture mechanism. In the HS car\-bon/e\-po\-xy material,
the transverse failure strain is approximately equal to 0.6\,\%,
while the longitudinal one is equal to 1.8\,\%. This type of fracture
generally has no direct effects on the fracture of the laminate, and
this approach therefore seems to be too conservative. Assuming that
the structure will be safe up to the occurrence of the first longitudinal
or shear failure, a more realistic torque resistance limit can be
obtained \citep{HochardCST05,Hochard09}. This limit can be computed
quite simply with a maximum stress criterion:}{\small \par}
{\small 
\begin{gather}
-X'\leq\sigma_{11}\leq X\quad;\quad\left|\sigma_{12}\right|\leq s
\end{gather}
}{\small \par}
{\small The comparisons made in Table~\ref{tab:failure-torque} between
the results obtained with these criteria and the experimental data
confirm the validity of this approach. In the case of tubes Nos.~1
and 2, the values obtained with the maximum stress criterion showed
better agreement with the experimental data than those obtained using
the Tsai-Wu criterion. The assumption involving the presence of a
null coupling mechanism in the case of unsymmetrical laminates was
also found to be true. The Tsai-Wu values could be improved by taking
a greater transverse tensile strength $Y$. }{\small \par}

\begin{table*}[t]
\caption{{\footnotesize Comparison between torque resistance calculations on
various short BE tubes \label{tab:failure-torque}}}

\begin{centering}
\begin{tabular}{lcccc}
\hline 
{\footnotesize Nos.} &  & {\footnotesize 1} & {\footnotesize 2} & {\footnotesize 3}\tabularnewline
\hline 
{\footnotesize Stacking sequence (from inner to outer radius)} & \textdegree{} & {\footnotesize {[}90,45,-45,90{]}} & {\footnotesize {[}90,45,-45,0$_{6}$,90{]}} & {\footnotesize {[}90,0$_{2}$,90{]}}\tabularnewline
\hline 
{\footnotesize Outer radius $\times$ length} & {\footnotesize mm} & {\footnotesize 25.4 $\times$50.8 } & {\footnotesize 63.5$\times$305} & {\footnotesize 25.4$\times$50.8}\tabularnewline
\hline 
{\footnotesize Experimental \citep{Zinberg70}} & {\footnotesize N.m} & {\footnotesize 581} & {\footnotesize 4689}\textcolor{black}{\footnotesize $^{a}$} & {\footnotesize 132}\tabularnewline
{\footnotesize Tsai-Wu criterion} & {\footnotesize N.m} & {\footnotesize 167 (-71\%)} & {\footnotesize 1605 (-66\%)} & {\footnotesize 130 (3\%)}\tabularnewline
{\footnotesize Tsai-Wu criterion with $\mathbf{B}=\mathbf{0}$} & {\footnotesize N.m} & {\footnotesize 313 (-46\%)} & {\footnotesize 2613 (-44\%)} & {\footnotesize 130 (3\%)}\tabularnewline
{\footnotesize Maximum stress in fibre and shear directions} & {\footnotesize N.m} & {\footnotesize 517 (-10\%)} & {\footnotesize 1610 (-66\%)} & {\footnotesize 130 (3\%)}\tabularnewline
{\footnotesize Maximum stress in fibre and shear directions with $\mathbf{B}=\mathbf{0}$} & {\footnotesize N.m} & {\footnotesize 585 (2\%)} & {\footnotesize 4880 (4\%)} & {\footnotesize 130 (3\%)}\tabularnewline
\hline 
{\footnotesize Buckling torque computed with \citet{Hayashi75} criterion} & {\footnotesize N.m} & {\footnotesize 1049} & {\footnotesize 13\,016} & {\footnotesize 1547}\tabularnewline
\hline 
\end{tabular}
\par\end{centering}

\centering{}\textcolor{black}{\footnotesize $^{a}$~Mean value of
two specimen tests.}
\end{table*}

\subsection{Torsional buckling analysis}

{\small Finite element methods are those most frequently employed
to predict torsional buckling. However, an alternative method is presented
here, which requires less computing time, and consists in solving
the buckling shell problem in the case of orthotropic circular cylinders,
using Flügge's buckling shell theory \citep{Flugge73}. The laminate
theory is included in the shell equations, as established in \citep{Bert95}.
This gives the Eqs.~(\ref{eq:buckling1}-\ref{eq:buckling3}) (see
\ref{sec:Torsional-buckling-equations}). Since we are dealing here
with very long shafts, it is possible to neglect the boundary condition
effects. In this case, a simplified displacement field presented by
Flügge can be used:
\begin{gather}
u=a\sin(h\varphi+\frac{p\pi x}{l}),\, v=b\sin(h\varphi+\frac{p\pi x}{l}),\, w=c\cos(h\varphi+\frac{p\pi x}{l})
\end{gather}
where $(u,v,w)$ is the displacement field of the middle-surface of
the cylinder, $h$ is the number of half-waves along the cylinder's
circumference and $p$ is the number of half-waves along the axis
of the cylinder with a fictive length $l$. When this displacement
field is applied to the shell equations, a classical eigenvalue problem
is obtained:}{\small \par}
{\small 
\begin{gather}
\mathbf{K}.\mathbf{U}=\mathbf{0}\quad\text{ with }\quad\mathbf{U}=\begin{bmatrix}a\\
b\\
c
\end{bmatrix}
\end{gather}
where $\mathbf{K}$ is the stiffness matrix (the elements of $\mathbf{K}$
are given in }\linebreak{}
{\small \ref{sec:Torsional-buckling-equations}). A non-trivial solution
exists when the determinant of $\mathbf{K}$ is null.}{\small \par}

{\small The numerical method used here consists in finding the minimum
value of the buckling torque $T_{\mathrm{buck}}$ which cancels the
determinant among all the values of $h\in\mathbb{N}^{*}$ and $p\in\mathbb{R}^{*+}$.
The computing time was reduced as follows. First, we have observed
that the minimum value of the buckling torque is always obtained at
$h=2$. Secondly, instead of searching for the value of $p$ between
$0$ and $+\infty$, this unknown can be found by searching around
the value of $l\left(48t^{2}/12r^{2}\right)^{1/4}/\pi r$ obtained
by \citet{Flugge73} in the case of isotropic material. Thirdly, the
search for the buckling torque is conducted around the value of $T_{\mathrm{buck}}$,
which can be obtained with an analytic criterion such as Hayashi's
criterion \citep{Hayashi75}:
\begin{gather}
T_{\mathrm{buck}}=11\sqrt{r}\left(A_{11}-\frac{A_{12}^{2}}{A_{22}}\right)^{1/4}D_{22}^{3/4}
\end{gather}
}{\small \par}

{\small It is worth noting that this criterion, like other classical
criteria, does not account for the coupling mechanism involved in
unsymmetrical laminates.}{\small \par}

{\small The shell method was first tested on off-axis stacking sequences.
The buckling torque obtained are presented in Table~\ref{tab:buckling}
and compared with those obtained with the finite element method, the
above criterion and by \citet{Bert95}. The results obtained with
the finite element method using ABAQUS (s4 elements) \citep{Abaqus01},
which were previously validated in \citep{MontagnierPhd} based on
experimental results obtained by \citet{Bauchau88} in the case of
short tube, are taken as reference values. \citeauthor{Bert95} buckling
theory is based on the Sanders shell theory and take the boundary
conditions into account. In the table, the results obtained by \citeauthor{Bert95}
correlate perfectly well with the finite element computations. Those
obtained with the method presented here show higher differences which
is explain by the too low length-to-diameter ratio ($l/(2r_{\mrm})\approx20$).
The results show the highest differences at the extremum cases (0\textdegree{}
and 90\textdegree{}) which yet are symmetric, hovewer, the method
is conservative. The error obtained with the Hayashi criterion can
reach 82\%.}{\small \par}

{\small The shell method was then tested on unsymmetrical stacking
sequences for higher length-to-diameter ratio ($l/(2r_{\mrm})=50$)
in Table~\ref{tab:buckling2}. All the tubes presented in the table
are of the same size and the laminates all have the same thickness.
Buckling torque was computed in the positive and then in the negative
direction. The table shows that the Hayashi criterion overestimates
the buckling torque, especially in the largest unsymmetrical laminates
(Nos. 9-12) by up to 40\%. The results obtained with shell theory
show good agreement with the finite element calculations, giving a
conservative estimate on the whole. The largest errors amounted to
only 8\% and the mean error was 4\%. }{\small \par}

\begin{table*}
\begin{centering}
\begin{tabular}{ccccccccc}
\hline 
{\footnotesize Ply orientation angle} & {\footnotesize \textdegree{}} & {\footnotesize 0} & {\footnotesize 15} & {\footnotesize 30} & {\footnotesize 45} & {\footnotesize 60} & {\footnotesize 75} & {\footnotesize 90}\tabularnewline
\hline 
{\footnotesize Abaqus} & {\footnotesize N.m} & {\footnotesize 1 489} & {\footnotesize 974} & {\footnotesize 1 121} & {\footnotesize 1 769} & {\footnotesize 2 587} & {\footnotesize 3 131} & {\footnotesize 3 278}\tabularnewline
\hline 
{\footnotesize Present work} & {\footnotesize N.m (\%)} & {\footnotesize 966 (-35)} & {\footnotesize 755 (-22)} & {\footnotesize 979 (-13)} & {\footnotesize 1\,647 (-7)} & {\footnotesize 2\,445 (-5)} & {\footnotesize 2\,957 (-6)} & {\footnotesize 2\,835 (-14)}\tabularnewline
{\footnotesize Bert \& Kim} & {\footnotesize N.m (\%)} & {\footnotesize 1\,587 (7)} & {\footnotesize 974 (0)} & {\footnotesize 1\,126 (0)} & {\footnotesize 1\,790 (1)} & {\footnotesize 2\,617 (1)} & {\footnotesize 3\,156 (1) } & {\footnotesize 3\,016 (-8)}\tabularnewline
{\footnotesize Hayashi criterion} & {\footnotesize N.m (\%)} & {\footnotesize 1\,887 (27)} & {\footnotesize 1\,776 (82)} & {\footnotesize 1\,607 (43)} & {\footnotesize 1\,648 (-7)} & {\footnotesize 2\,216 (-14)} & {\footnotesize 3\,925 (-25)} & {\footnotesize 3\,365 (3)}\tabularnewline
\hline 
\end{tabular}
\par\end{centering}

\caption{{\footnotesize Comparison between buckling torque calculations on
off-axis BE drive shafts ($l=2.47$~m, $r_{\mathrm{m}}=62.85$~mm
and $t_{\srm}=1.32$~mm)\label{tab:buckling}}}
\end{table*}

\begin{table*}[t]
\caption{{\footnotesize Comparison between buckling torque calculations on
unsymmetrical CFRP drive shafts ($l=4$~m, $r_{\mathrm{m}}=40$~mm,
$E_{11}=134$~GPa, $E_{22}=8.5$~GPa, $E_{66}=E_{55}=4.6$GPa, $E_{44}=4.0$~GPa,
$E_{12}=0.29$ and $t_{\srm}=1.067$~mm)\label{tab:buckling2}}}

\begin{centering}
\begin{tabular}{lllcccccccc}
\hline 
\multicolumn{2}{l}{{\footnotesize Laminate}} &  & \multicolumn{2}{l}{{\footnotesize ABAQUS}} &  & \multicolumn{2}{l}{{\footnotesize Present}} &  & \multicolumn{2}{l}{{\footnotesize Hayashi}}\tabularnewline
 &  &  & \multicolumn{2}{c}{} &  & \multicolumn{2}{l}{{\footnotesize work}} &  & \multicolumn{2}{l}{{\footnotesize criterion}}\tabularnewline
\cline{4-5}\cline{7-8}\cline{10-11}{\footnotesize Nos.} &  &  & {\footnotesize Mesh$^{a}$} & {\footnotesize Nm} &  & {\footnotesize Nm} & {\footnotesize \%} &  & {\footnotesize Nm} & {\footnotesize \%}\tabularnewline
\hline 
{\footnotesize 1} & {\footnotesize {[}15,-15{]}$_{4}$} &  & {\footnotesize 60-150} & {\footnotesize 210} &  & {\footnotesize 193} & {\footnotesize -8} &  & {\footnotesize 222} & {\footnotesize 6 }\tabularnewline
{\footnotesize 2} & {\footnotesize {[}-15,15{]}$_{4}$} &  & {\footnotesize 60-150} & {\footnotesize 214} &  & {\footnotesize 197} & {\footnotesize -8} &  & {\footnotesize 222} & {\footnotesize 4 }\tabularnewline
{\footnotesize 3} & {\footnotesize {[}30,-30{]}$_{4}$} &  & {\footnotesize 60-150} & {\footnotesize 263} &  & {\footnotesize 254} & {\footnotesize -4} &  & {\footnotesize 283} & {\footnotesize 8 }\tabularnewline
{\footnotesize 4} & {\footnotesize {[}-30,30{]}$_{4}$} &  & {\footnotesize 60-150} & {\footnotesize 268} &  & {\footnotesize 259} & {\footnotesize -3} &  & {\footnotesize 283} & {\footnotesize 6}\tabularnewline
{\footnotesize 5} & {\footnotesize {[}45,-45{]}$_{4}$} &  & {\footnotesize 60-150} & {\footnotesize 385} &  & {\footnotesize 383} & {\footnotesize -1} &  & {\footnotesize 419} & {\footnotesize 9}\tabularnewline
{\footnotesize 6} & {\footnotesize {[}-45,45{]}$_{4}$} &  & {\footnotesize 60-150} & {\footnotesize 385} &  & {\footnotesize 382} & {\footnotesize -1} &  & {\footnotesize 419} & {\footnotesize 9}\tabularnewline
{\footnotesize 7} & {\footnotesize {[}0$_{2}$,45,-45,45,-45,0$_{2}${]}} &  & {\footnotesize 60-150} & {\footnotesize 230} &  & {\footnotesize 218} & {\footnotesize -5} &  & {\footnotesize 252} & {\footnotesize 10 }\tabularnewline
{\footnotesize 8} & {\footnotesize {[}0$_{2}$,-45,45,-45,45,0$_{2}${]}} &  & {\footnotesize 60-150} & {\footnotesize 219} &  & {\footnotesize 208} & {\footnotesize -5} &  & {\footnotesize 252} & {\footnotesize 15 }\tabularnewline
{\footnotesize 9} & {\footnotesize {[}0$_{2}$,45,0,-45,0,45,-45{]}} &  & {\footnotesize 30-100} & {\footnotesize 358} &  & {\footnotesize 342} & {\footnotesize -4} &  & {\footnotesize 420} & {\footnotesize 17 }\tabularnewline
{\footnotesize 10} & {\footnotesize {[}0$_{2}$,-45,0,45,0,-45,45{]}} &  & {\footnotesize 30-100} & {\footnotesize 329} &  & {\footnotesize 315} & {\footnotesize -4} &  & {\footnotesize 420} & {\footnotesize 28 }\tabularnewline
{\footnotesize 11} & {\footnotesize {[}0$_{2}$,45,0$_{2}$,-45,45,-45{]}} &  & {\footnotesize 30-100} & {\footnotesize 355} &  & {\footnotesize 340} & {\footnotesize -4} &  & {\footnotesize 440} & {\footnotesize 24 }\tabularnewline
{\footnotesize 12} & {\footnotesize {[}0$_{2}$,-45,0$_{2}$,45,-45,45{]}} &  & {\footnotesize 30-100} & {\footnotesize 313} &  & {\footnotesize 300} & {\footnotesize -4} &  & {\footnotesize 440} & {\footnotesize 41 }\tabularnewline
{\footnotesize 13} & {\footnotesize {[}-45,-15,15,45,15,-15,-45,45{]}} &  & {\footnotesize 60-150} & {\footnotesize 389} &  & {\footnotesize 375} & {\footnotesize -4} &  & {\footnotesize 493} & {\footnotesize 27}\tabularnewline
{\footnotesize 14} & {\footnotesize {[}45,15,-15,-45,-15,15,45,-45{]}} &  & {\footnotesize 60-150} & {\footnotesize 439} &  & {\footnotesize 449} & {\footnotesize 2} &  & {\footnotesize 493} & {\footnotesize 12 }\tabularnewline
{\footnotesize 15} & {\footnotesize {[}15,-15,-45,-15,15,45,15,-15{]}} &  & {\footnotesize 60-150} & {\footnotesize 219} &  & {\footnotesize 206} & {\footnotesize -6} &  & {\footnotesize 265} & {\footnotesize 21 }\tabularnewline
{\footnotesize 16} & {\footnotesize {[}-15,15,45,15,-15,-45,-15,15{]}} &  & {\footnotesize 60-150} & {\footnotesize 241} &  & {\footnotesize 226} & {\footnotesize -6} &  & {\footnotesize 265} & {\footnotesize 10}\tabularnewline
\hline 
\end{tabular}
\par\end{centering}

\centering{}\textcolor{black}{\footnotesize $^{a}$~Number of circumferential
elements - number of lengthwise elements.}
\end{table*}

\subsection{Driveline mass}

{\small The driveline is composed of shafts and intermediate supports,
which include bearings, fittings, and supports. The intermediate support
mass $m_{\brm}$ can be computed with an empirical equation from \citet{Lim86}:
\begin{gather}
m_{\brm}=17.1288\left(\frac{P_{\mathrm{dv}}}{\Omega_{\mathrm{nom}}}\right)^{0.69}\label{eq:masse_paliers}
\end{gather}
}{\small \par}

{\small where $P_{\mathrm{dv}}$ is the power transmitted with the
driveline (in W) and $\Omega_{\mathrm{nom}}$ is the nominal spin
speed (in rev\,/\,min). The driveline's mass can then be computed
using the following expression: }{\small \par}

{\small 
\begin{gather}
m_{\mathrm{dv}}=N_{\srm}\times m_{\srm}+N_{\brm}\times m_{\brm}\quad\mbox{ with }\quad m_{\srm}=\rho Sl
\end{gather}
}{\small \par}

{\small where $N_{\srm}$ is the number of shafts and $N_{\brm}$
is the number of intermediate supports.}{\small \par}

\section{Shaft optimisation using a genetic algorithm}

{\small The principles underlying the GA algorithm are the same as
those on which Darwin's theory of evolution was based. At the beginning,
a randomly created population is evaluated with a fitness function.
The result gives the fitness of each individual. Starting with this
information, the new generation of the population can be deduced using
selection, crossover and mutation operators. This process is iterated
up to convergence. }{\small \par}

{\small The main risk with this stochastic method is that of not obtaining
the optimum solution. In particular, GA may tend to converge on local
optima and may not be able to cross these attracting points. Another
weakness of the method is the large amount of fitness function calculations
required. This means that the evaluation procedure must not be too
time-consuming. }{\small \par}

\subsection{Individual}

{\small An individual in this driveline optimisation procedure consists
of the medium diameter $r_{\mathrm{m}}$ (which can be fixed or otherwise),
the bearing stiffness $k_{\erm}$ (fixed or not), the nominal spin
speed $\Omega_{\mathrm{nom}}$ and the stacking sequence with various
materials, symmetric or otherwise, as in the following example: $[\alpha_{1}^{\mathrm{mat}_{1}}\times n_{1},\,...,\alpha_{j}^{\mathrm{mat}_{j}}\times n_{j},\,...,\alpha_{q}^{\mathrm{mat}_{q}}\times n_{q}]$
where $\alpha_{j}$, $n_{j}$ and $\mathrm{mat}_{j}$ are the orientation,
the number of plies and the material of which the ply $j$ is made,
respectively. Under supercritical conditions, the stiffness of the
bearings is a necessary optimisation variable because it appears in
both the rigid mode frequencies Eq.~(\ref{eq:parametres}) and the
stability criterion Eq.~(\ref{eq:lim.hystplus}). The shaft length
corresponds to the driveline length divided by the number of shafts. }{\small \par}

{\small There are several possible ways of modelling the chromosomes
of individuals. It is proposed here to fix the number of possible
orientations in the stacking sequence, denoted $q$. This sets the
size of the chromosome in the case of a particular optimisation process,
which simplifies the crossover operations. Individuals are classically
represented by an array of binary numbers. The orientation $\alpha_{j}$
can be written with 2 or 3 bits, standing for the sets \{-45, 0, 45,
90\} or \{-67.5, -45, -22.5, 0, 22.5, 45, 67.5, 90\} (in degree units),
respectively, which correspond to realistic prepreg hand lay-up orientations.
The quantity $n_{j}$ is written with 2 or 3 bits corresponding to
the sets \{1,2,3,4\} and \{1,2,3,4,5,6,7,8\}, respectively. The material
$\mathrm{mat}_{j}$ is written with one bit to take advantage of both
HM and HS car\-bon/e\-po\-xy, or metal and HM car\-bon/e\-po\-xy,
for example. Lastly, $k_{\erm}$ and $r_{\mathrm{m}}$ are bounded
and generally encoded with 3 bits. For example, a shaft with the following
stacking sequence $[45{}_{2}^{\mathrm{mat}_{1}},0{}_{3}^{\mathrm{mat}_{2}}]$
(i.e. $q=2$), with $r_{\mathrm{m}}=$52\,mm and $\Omega_{\mathrm{nom}}=$4000\,rev\,/\,min,
with the bearing stiffness fixed and where $\alpha_{j}$ and $n_{j}$
are encoded with 2 bits, $r_{\mathrm{m}}$ with 4 bits, $\Omega_{\mathrm{nom}}$
with 3 bits and $\mathrm{mat}_{j}$ with 1 bit, is defined by the
following chromosome:
\begin{gather*}
\underbrace{\begin{array}{|c|c|}
\hline 1 & 0\\\hline \end{array}}_{\alpha_{1}=45{^\circ}}\underbrace{\begin{array}{|c|c|}
\hline 0 & 1\\\hline \end{array}}_{n_{1}=2}\underbrace{\begin{array}{|c|}
\hline 0\\\hline \end{array}}_{\mathrm{mat}_{1}}\underbrace{\begin{array}{|c|c|}
\hline 0 & 1\\\hline \end{array}}_{\alpha_{2}=0{^\circ}}\underbrace{\begin{array}{|c|c|}
\hline 1 & 0\\\hline \end{array}}_{n_{2}=3}\underbrace{\begin{array}{|c|}
\hline 1\\\hline \end{array}}_{\mathrm{mat}_{2}}...\\
...\underbrace{\begin{array}{|c|c|c|c|}
\hline 1 & 0 & 1 & 1\\\hline \end{array}}_{r_{\mathrm{m}}=52\in[30,60]}\underbrace{\begin{array}{|c|c|c|}
\hline 0 & 1 & 1\\\hline \end{array}}_{\Omega_{\mathrm{nom}}=4000\in[3700,4400]}
\end{gather*}
}{\small \par}

{\small The string length is therefore simply $(\mathrm{bit}_{\alpha}+\mathrm{bit}_{n}+\mathrm{bit}_{\mathrm{mat}})\times q+\mathrm{bit}_{k_{\erm}}+\mathrm{bit}_{r_{\mathrm{m}}}+\mathrm{bit}_{\Omega_{\mathrm{nom}}}$.}{\small \par}

\subsection{Constraints and fitness}

{\small The mass is the optimised value generally used in driveline
problems \citep{Lim86,Gubran02a}. In this paper, part of the fitness
function is equal to the inverse of the mass of one shaft. The other
part depends on the strength, buckling and dynamic constraints previously
investigated ($n\in\mathbb{N}^{*}$):
\begin{gather}
g_{1}=\frac{K_{\mathrm{str}}T_{\mathrm{str}}}{T_{\mathrm{nom}}}-1\geq0\mbox{\quad with\quad}K_{\mathrm{str}}\leq1\label{eq:contraint_tsai}\\
g_{2}=\frac{K_{\mathrm{buck}}T_{\mathrm{buck}}}{T_{\mathrm{nom}}}-1\geq0\mbox{\quad with\quad}K_{\mathrm{buck}}\leq1\label{eq:contraint_buck}\\
g_{3}=\frac{t_{\mathrm{\srm}}}{t_{\mathrm{\srm\, min}}}-1\geq0\label{eq:contrainte_radius}\\
g_{4n}=1-\frac{K_{\mathrm{t\, inf}\, n}\Omega_{n}}{\Omega_{\mathrm{nom}}}\geq0\mbox{\quad with\quad}K_{\mathrm{t\, inf}\, n}\geq1\label{eq:contraint_tors1}\\
g_{5n}=\frac{K_{\mathrm{t\, sup}\, n}\Omega_{n}}{\Omega_{\mathrm{nom}}}-1\geq0\mbox{\quad with\quad}K_{\mathrm{t\, sup}\, n}\leq1\label{eq:contraint_tors2}
\end{gather}
and in the subcritical case
\begin{gather}
g_{6}=\frac{K_{\mathrm{f\, sup}\,1}\omega_{\crm1}}{\Omega_{\mathrm{nom}}}-1>0\mbox{\quad with\quad}K_{\mathrm{f\, sup}\,1}\leq1\label{eq:contraint_flex_sub}
\end{gather}
or in the supercritical case
\begin{gather}
g_{7n}=1-\frac{K_{\mathrm{f\, inf}\, n}\omega_{\crm n}}{\Omega_{\mathrm{nom}}}\geq0\mbox{\quad with\quad}K_{\mathrm{f\, inf}\, n}\geq1\label{eq:contraint_flex_sup1}\\
g_{8n}=\frac{K_{\mathrm{f\, sup}\, n}\omega_{\crm n}}{\Omega_{\mathrm{nom}}}-1\geq0\mbox{\quad with\quad}K_{\mathrm{f\, sup}\, n}\leq1\label{eq:contraint_flex_sup2}\\
g_{9}=\frac{K_{\mathrm{th}}\omega_{\ths}}{\Omega_{\mathrm{nom}}}-1\geq0\mbox{\quad with\quad}K_{\mathrm{th}}\leq1\label{eq:contraint_flex_sup3}
\end{gather}
where $g_{i}$ and $K_{...}$ are constraint functions and reserve
factors, respectively. Eq.~(\ref{eq:contraint_tsai}) corresponds
to the torsional strength constraint, which requires that the torque
computed with the strength criterion multiplied by the reserve factor
is smaller than the torque required. Eq.~(\ref{eq:contraint_buck})
is the same equation but for the torsional buckling. The other equations
are those giving the dynamic constraints. Eqs.~(\ref{eq:contraint_tors1}-\ref{eq:contraint_tors2})
correspond to the positioning of the nominal spin speed between torsional
modal frequencies. As regards the bending modes, the constraints depend
on whether the subcritical or supercritical case applies. In the first
case, Eq.~(\ref{eq:contraint_flex_sub}) corresponds to the subcritical
assumption, i.e. the nominal spin speed multiplied by the reserve
factor must be smaller than the first critical speed. In the supercritical
case, Eqs.~(\ref{eq:contraint_flex_sup1}-\ref{eq:contraint_flex_sup2})
correspond to the positioning of the nominal spin speed between the
flexural critical speeds, and Eq.~(\ref{eq:contraint_flex_sup3})
corresponds to the stability constraint.  }{\small \par}

{\small GAs cannot account directly for constraint functions. This
problem can be overcome by using a penalisation method consisting
in deteriorating the quality of an individual that violates one or
more constraints, by decreasing the fitness function. The fitness
function used for mass minimisation purposes can be written in the
following general form: }{\small \par}

{\small 
\begin{gather}
f=\frac{1}{m_{\mathrm{s}}}+\sum_{j}\gamma_{j}\min(0,g_{j})
\end{gather}
}{\small \par}

{\small where $\gamma_{j}$ are the penalisation factors. The reserve
factors and penalisation factors are given in Table~\ref{tab:ag_rf_and_pf}.}{\small \par}

\begin{table*}[!t]
\caption{{\footnotesize Reserve factors and penalisation factors \label{tab:ag_rf_and_pf}}}

\centering{}%
\begin{tabular}{cccccccccc}
\hline 
{\footnotesize $K_{\mathrm{str}}$} & {\footnotesize $K_{\mathrm{buck}}$} & {\footnotesize $K_{\mathrm{t\, sup}\, n}$} & {\footnotesize $K_{\mathrm{t\, inf}\, n}$} & {\footnotesize $K_{\mathrm{f\, sup}\, n}$} & {\footnotesize $K_{\mathrm{f\, inf}\, n}$} & {\footnotesize $K_{\mathrm{th}}$} & {\footnotesize $\gamma_{1,2}$} & {\footnotesize $\gamma_{3}$} & {\footnotesize $\gamma_{j\notin\{1,2,3\}}$}\tabularnewline
\hline 
{\footnotesize 0.44} & {\footnotesize 0.44} & {\footnotesize 0.83} & {\footnotesize 1.15} & {\footnotesize 0.8} & {\footnotesize 1.2} & {\footnotesize 0.8} & {\footnotesize 2} & {\footnotesize 6} & {\footnotesize 4}\tabularnewline
\hline 
\end{tabular}
\end{table*}

\subsection{The genetic algorithm method}

\subsubsection{Initialisation}

{\small The algorithm is initialised by randomly generating a population
of 300-600 individuals. The number depends on the size of the problem.}{\small \par}

\subsubsection{Elitism}

{\small After evaluating the population with the fitness function,
the two fittest individuals, which are also called the elites, are
selected and kept for the next generation. }{\small \par}

\subsubsection{Scaling, selection and crossover}

{\small With the progression of the GA, the fitness of all the individuals
tends to converge on that of the fittest ones. This slows down the
progress of the algorithm. A windowing method \citep{Goldberg89}
is used here, whereby the fitness of the lowest ranking individual
is subtracted from the fitness of each individual. Two parents are
then selected, based on their scaled fitness values and a multi-point
crossover operation is performed. The cutting point is selected randomly.
This operation gives two children, forming the next generation.}{\small \par}

\subsubsection{Mutation}

{\small The mutation consists in randomly modifying the bits of the
chromosomes. The probability of occurrence of the mutation must be
very high to obtain a highly diverse population. But if the mutation
process is too strong, the algorithm may not converge on the optimum
fitness. Note that elites are not subject to mutations. After the
mutation, the process is restarted at the elitism stage until the
maximum fitness function is reached. The search parameters in the
GA are given in Table~\ref{tab:ag_parameter}.}{\small \par}

\begin{table}
\caption{{\footnotesize Search parameters of the genetic algorithm \label{tab:ag_parameter}}}

\centering{}%
\begin{tabular}{p{57mm}>{\raggedleft}p{16mm}}
\hline 
{\footnotesize Population size } & {\footnotesize 300-600}\tabularnewline
{\footnotesize Chromosome length } & {\footnotesize 24-34}\tabularnewline
{\footnotesize Crossover probability } & {\footnotesize 90\%}\tabularnewline
{\footnotesize Mutation probability } & {\footnotesize 10\%}\tabularnewline
{\footnotesize Number of generation } & {\footnotesize 150-40000}\tabularnewline
\hline 
\end{tabular}
\end{table}

\section{Case study}

{\small The helicopter tail rotor driveline presented by \citet{Zinberg70}
is investigated with the GA. The original driveline, having a total
length of 7.41\,m, which is assumed to transmit a power of 447.4\,kW,
is composed of five subcritical aluminium alloy tubes and four intermediate
supports. \citeauthor{Zinberg70} suggested replacing the conventional
driveline by three subcritical composite shafts consisting of BE material.
The properties of this composite shaft are compared with those of
the aluminium one in Table~\ref{tab:compare_shaft-sub}. Note that
the Zinberg shaft was obtained only on the basis of physical considerations. }{\small \par}

{\small In line with \citet{Lim86}, who studied the same driveline
case, the mass moment of inertia of the main gearing and tail rotor
are assumed to be equal to 0.94\,kg\,m$^{2}$ and 3.76\,kg\,m$^{2}$,
respectively. To take the difference between the connections in the
metallic and composite shafts into account, a weight penalty of 1.5\,kg
per composite shaft is added here. }{\small \par}

\begin{table*}[t]
\caption{{\footnotesize Optimised composite tail rotor driveline under subcritical
conditions in comparison with the conventional aluminium driveline
\label{tab:compare_shaft-sub}}}

\begin{centering}
{\footnotesize }%
\begin{tabular}{llccccccc}
\hline 
 &  &  & {\footnotesize Conv.} & {\footnotesize Zinberg} & \multicolumn{4}{c}{{\footnotesize Optimised}}\tabularnewline
\cline{4-9} 
{\footnotesize Material} &  &  & {\footnotesize aluminium} & {\footnotesize BE\citep{Zinberg70}} & {\footnotesize BE} & {\footnotesize HM} & {\footnotesize HM}\textcolor{black}{\footnotesize $^{b}$} & {\footnotesize HS/HM}\tabularnewline
\hline 
{\footnotesize Number of tubes} &  & {\footnotesize -} & {\footnotesize 5} & \multicolumn{5}{c}{{\footnotesize 3}}\tabularnewline
\hline 
{\footnotesize String length} &  & {\footnotesize bit} & {\footnotesize -} & {\footnotesize -} & {\footnotesize 24} & {\footnotesize 24} & {\footnotesize 30} & {\footnotesize 30}\tabularnewline
\hline 
{\footnotesize Stacking sequence} &  &  &  & {\footnotesize {[}90,45,} & {\footnotesize {[}90$_{2}$,0$_{4}$,} & {\footnotesize {[}90,0$_{3}$,} & {\footnotesize {[}90,-22.5$_{2}$,} & {\footnotesize {[}$90^{\mathrm{HM}}$,$45^{\mathrm{HR}}$,}\tabularnewline
{\footnotesize (from inner to } &  & {\footnotesize \textdegree{}} & {\footnotesize -} & {\footnotesize -45,0$_{6}$,} & {\footnotesize -45,45,} & {\footnotesize 45,-45$_{2}$,} & {\footnotesize 22.5,-22.5,} & {\footnotesize $0_{4}^{\mathrm{HM}}$,$-45^{\mathrm{HR}}$,}\tabularnewline
{\footnotesize outer radius)} &  &  &  & {\footnotesize 90{]}} & {\footnotesize 90{]}} & {\footnotesize 45{]}} & {\footnotesize 22.5$_{2}$,-67.5{]}} & {\footnotesize $90^{\mathrm{HM}}${]} }\tabularnewline
\hline 
{\footnotesize Operating speed} & {\footnotesize $\Omega_{\mathrm{nom}}$} & {\footnotesize rev\,/\,min} & {\footnotesize 5\,540} & {\footnotesize 4\,320} & {\footnotesize 3\,800} & {\footnotesize 4\,800} & {\footnotesize 4\,600} & {\footnotesize 4\,400}\tabularnewline
{\footnotesize 1st critical speed} & {\footnotesize $\omega_{1}$} & {\footnotesize rev\,/\,min} & {\footnotesize 8\,887 } & {\footnotesize 5\,697}\textcolor{black}{\footnotesize $^{a}$} & {\footnotesize 4\,606} & {\footnotesize 5\,800} & {\footnotesize 5\,695} & {\footnotesize 5344}\tabularnewline
{\footnotesize 1st torsion mode} & {\footnotesize $\varpi_{1}$} & {\footnotesize rev\,/\,min} & {\footnotesize 2\,058}\textcolor{black}{\footnotesize $^{a}$} & \textcolor{black}{\footnotesize 1\,292$^{a}$} & {\footnotesize 1\,065} & {\footnotesize 1\,534} & {\footnotesize 1\,254} & {\footnotesize 635}\tabularnewline
{\footnotesize 2nd torsion mode} & {\footnotesize $\varpi_{2}$} & {\footnotesize rev\,/\,min} & {\footnotesize 65\,370}\textcolor{black}{\footnotesize $^{a}$} & {\footnotesize 35}\textcolor{black}{\footnotesize \,}{\footnotesize 318}\textcolor{black}{\footnotesize $^{a}$} & {\footnotesize 36\,428} & {\footnotesize 64\,965} & {\footnotesize 59\,599} & {\footnotesize 34\,510}\tabularnewline
\hline 
{\footnotesize Nominal torque} & {\footnotesize $T_{\mathrm{nom}}$} & {\footnotesize N\,m} & {\footnotesize 771} & {\footnotesize 989} & {\footnotesize 1\,124} & {\footnotesize 891} & {\footnotesize 929} & {\footnotesize 971}\tabularnewline
{\footnotesize Strength torque} & {\footnotesize $T_{\mathrm{str}}$} & {\footnotesize N\,m} & \textcolor{black}{\footnotesize 4\,925$^{a}$} & {\footnotesize 4\,880 }\textcolor{black}{\footnotesize $^{a}$} & {\footnotesize 3\,149} & {\footnotesize 2\,268} & {\footnotesize 2\,267} & {\footnotesize 3\,349}\tabularnewline
{\footnotesize Buckling torque} & {\footnotesize $T_{\mathrm{buck}}$} & {\footnotesize N\,m} & \textcolor{black}{\footnotesize 3\,090$^{a}$} & {\footnotesize 2\,671}\textcolor{black}{\footnotesize $^{a}$}{\footnotesize{} } & {\footnotesize 2\,645} & {\footnotesize 2\,108} & {\footnotesize 2\,105} & {\footnotesize 2\,206}\tabularnewline
\hline 
{\footnotesize Tube length} & {\footnotesize $l$} & {\footnotesize m} & {\footnotesize 1.482} & \multicolumn{5}{c}{{\footnotesize 2.470}}\tabularnewline
{\footnotesize Mean tube radius} & {\footnotesize $r_{\mathrm{m}}$} & {\footnotesize mm} & {\footnotesize 56.3} & {\footnotesize 62.84} & {\footnotesize 56} & {\footnotesize 54} & {\footnotesize 50} & {\footnotesize 46}\tabularnewline
{\footnotesize Tube thickness} & {\footnotesize $t_{\mathrm{s}}$} & {\footnotesize mm} & {\footnotesize 1.65} & {\footnotesize 1.321} & {\footnotesize 1.19} & {\footnotesize 1.00} & {\footnotesize 1.00} & {\footnotesize 1.00}\tabularnewline
\hline 
{\footnotesize Tubes weight} & {\footnotesize $N_{\srm}m_{\mathrm{s}}$} & {\footnotesize kg (\%)} & {\footnotesize 13.38} & {\footnotesize 8.16 (61)} & {\footnotesize 6.09 (46)} & {\footnotesize 4.26 (32)} & {\footnotesize 3.96 (30)} & {\footnotesize 3.57 (27)}\tabularnewline
{\footnotesize Supports weight} & {\footnotesize $N_{\brm}m_{\mathrm{b}}$} & {\footnotesize kg} & {\footnotesize 15.42} & {\footnotesize 7.71} & {\footnotesize 9.68} & {\footnotesize 8.24} & {\footnotesize 8.48} & {\footnotesize 8.75}\tabularnewline
{\footnotesize Weight penalty} &  & {\footnotesize kg} & {\footnotesize -} & {\footnotesize 4.5}\textcolor{black}{\footnotesize $^{a}$} & {\footnotesize 4.5} & {\footnotesize 4.5} & {\footnotesize 4.5} & {\footnotesize 4.5}\tabularnewline
{\footnotesize Total weight} & {\footnotesize $m_{\mathrm{dv}}$} & {\footnotesize kg} & {\footnotesize 28.80} & {\footnotesize 20.37}\textcolor{black}{\footnotesize $^{a}$} & {\footnotesize 20.27} & {\footnotesize 17} & {\footnotesize 16.95} & {\footnotesize 16.82}\tabularnewline
{\footnotesize Weight saving} &  & {\footnotesize kg (\%)} & {\footnotesize - } & {\footnotesize 8.4 (29)}\textcolor{black}{\footnotesize $^{a}$} & {\footnotesize 8.5 (30)} & {\footnotesize 11.8 (41)} & {\footnotesize 11.9 (41)} & {\footnotesize 12.0 (42)}\tabularnewline
\hline 
\end{tabular}
\par\end{centering}{\footnotesize \par}

\centering{}\textcolor{black}{\footnotesize $^{a}$~Value computed
with presented methods. $^{b}$~with }{\footnotesize $\mathrm{bit}_{\alpha}=3$}
\end{table*}

\subsection{Subcritical shaft optimisation}

\begin{figure*}[!t]
\begin{centering}
\subfloat[{{\footnotesize BE ($\mathrm{bit}_{\alpha}=2$, $\mathrm{bit}_{\mathrm{mat}}=0$,
$r_{\mathrm{m}}\in$ {[}0.05,~0.064{]}\,mm)\label{fig:sous-BE}}}]{\begin{centering}
\includegraphics[height=7.2cm]{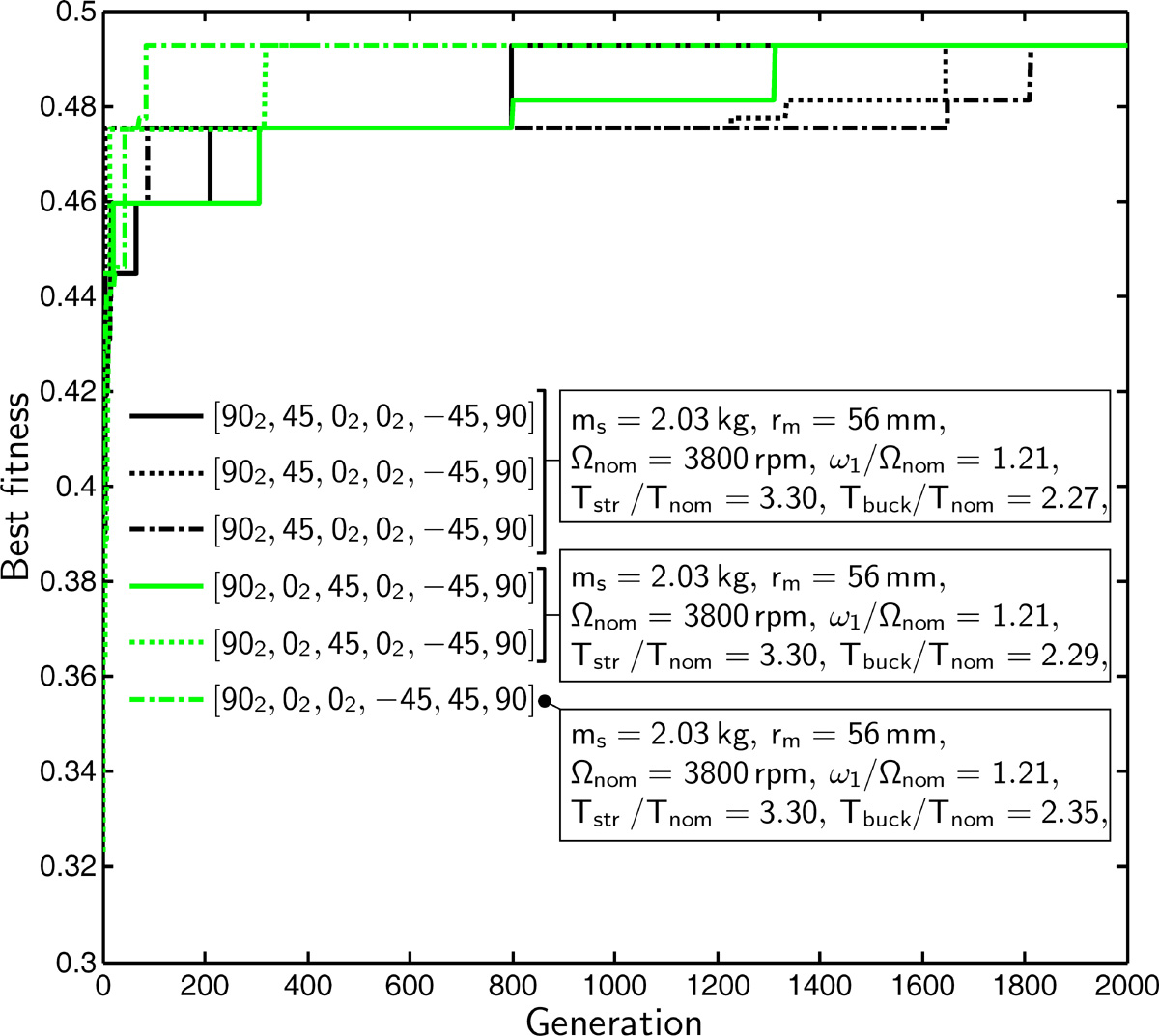}
\par\end{centering}

}\quad{}\subfloat[{{\footnotesize HM ($\mathrm{bit}_{\alpha}=2$, $\mathrm{bit}_{\mathrm{mat}}=0$,
$r_{\mathrm{m}}\in$ {[}0.05,~0.064{]}\,mm)\label{fig:sous-HM}}}]{\begin{centering}
\includegraphics[height=7.2cm]{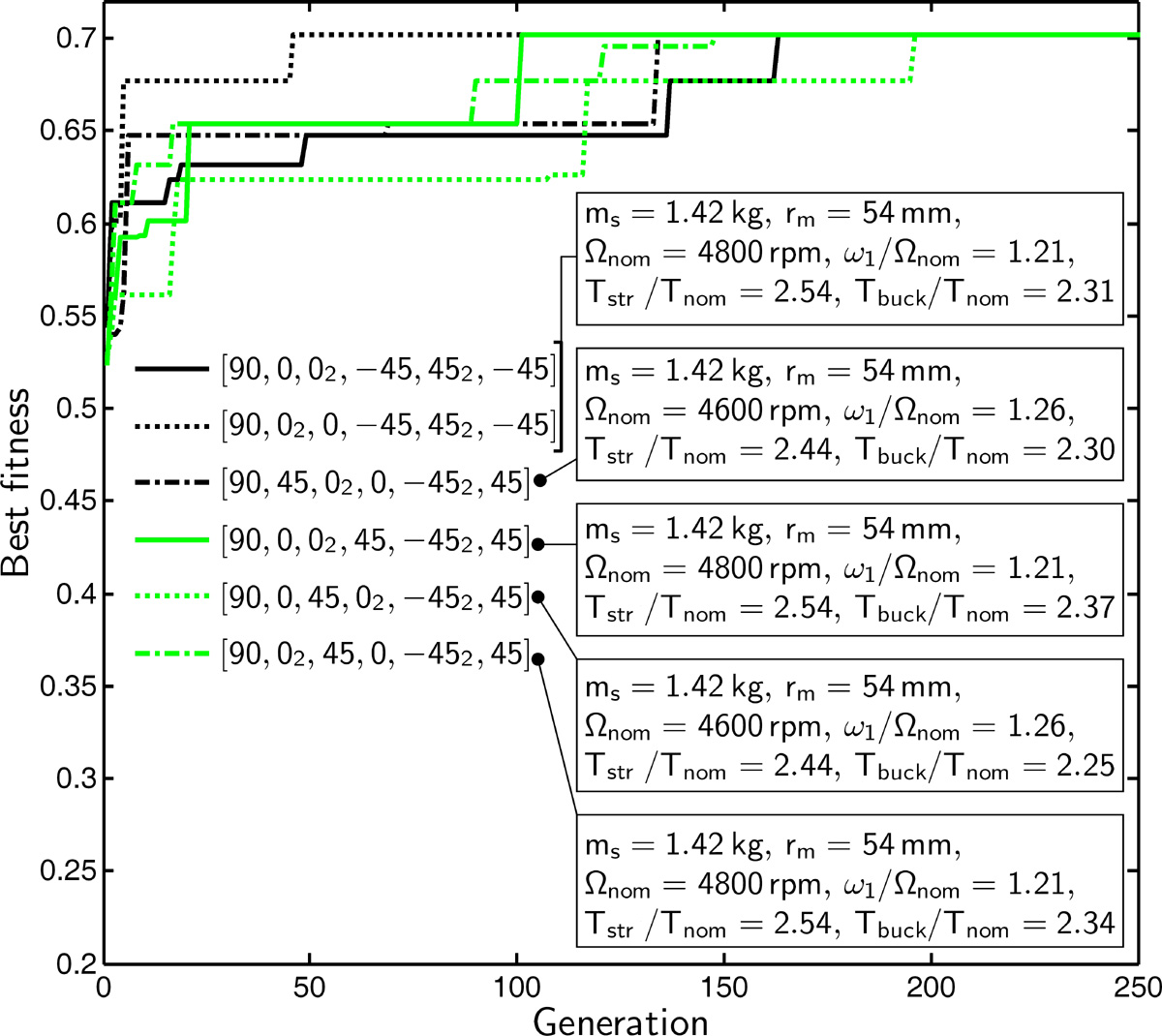}
\par\end{centering}

}
\par\end{centering}

\begin{centering}
\subfloat[{{\footnotesize HM ($\mathrm{bit}_{\alpha}=3$, $\mathrm{bit}_{\mathrm{mat}}=0$,
$r_{\mathrm{m}}\in$ {[}0.04,~0.054{]}\,mm)\label{fig:sous-HM-1}}}]{\begin{centering}
\includegraphics[height=7.2cm]{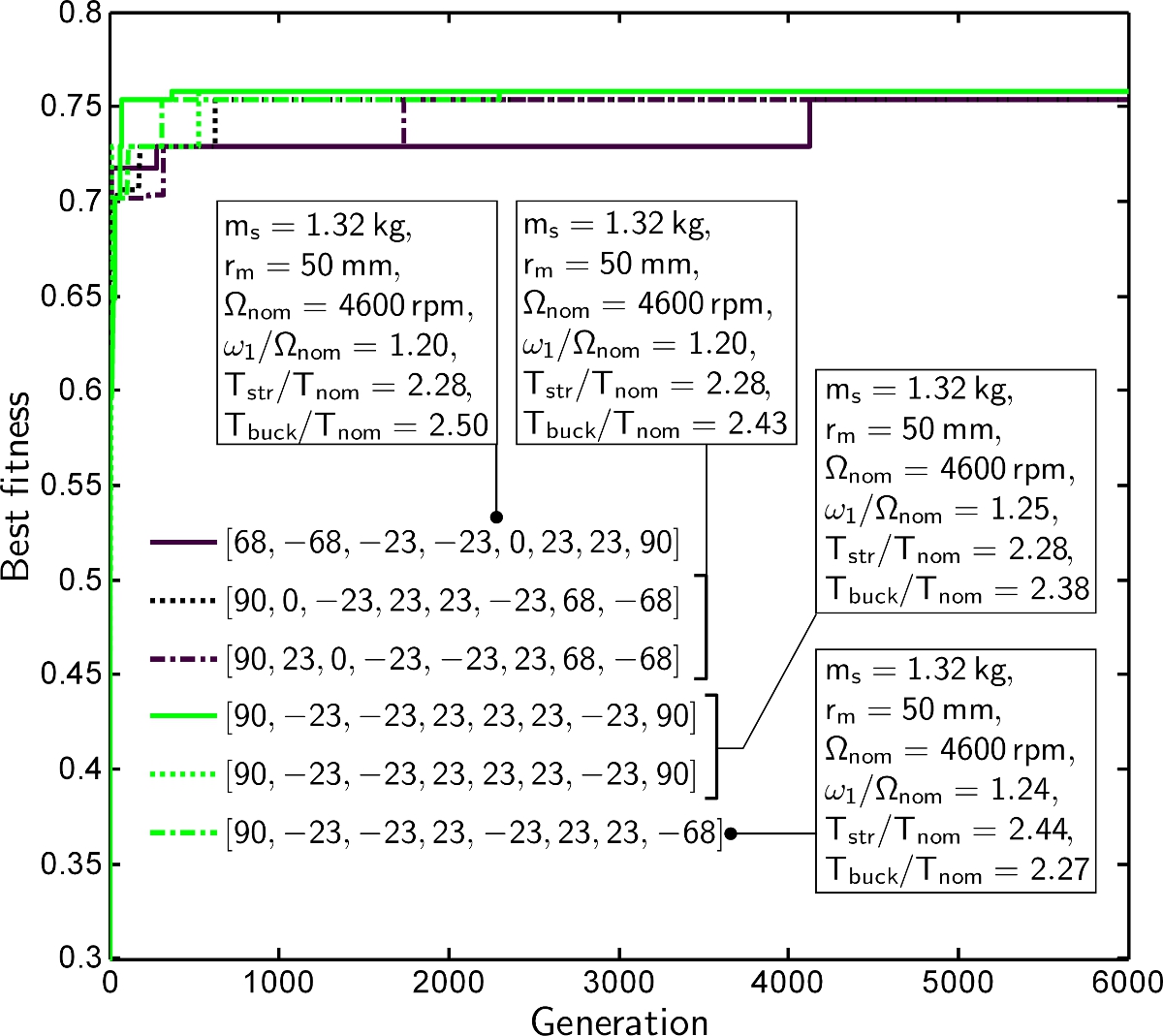}
\par\end{centering}

}\quad{}\subfloat[{{\footnotesize HM/HS ($\mathrm{bit}_{\alpha}=2$, $\mathrm{bit}_{\mathrm{mat}}=1$,
$r_{\mathrm{m}}\in$ {[}0.04,~0.054{]}\,mm) \label{fig:sous-HM-HR}}}]{\begin{centering}
\includegraphics[height=7.2cm]{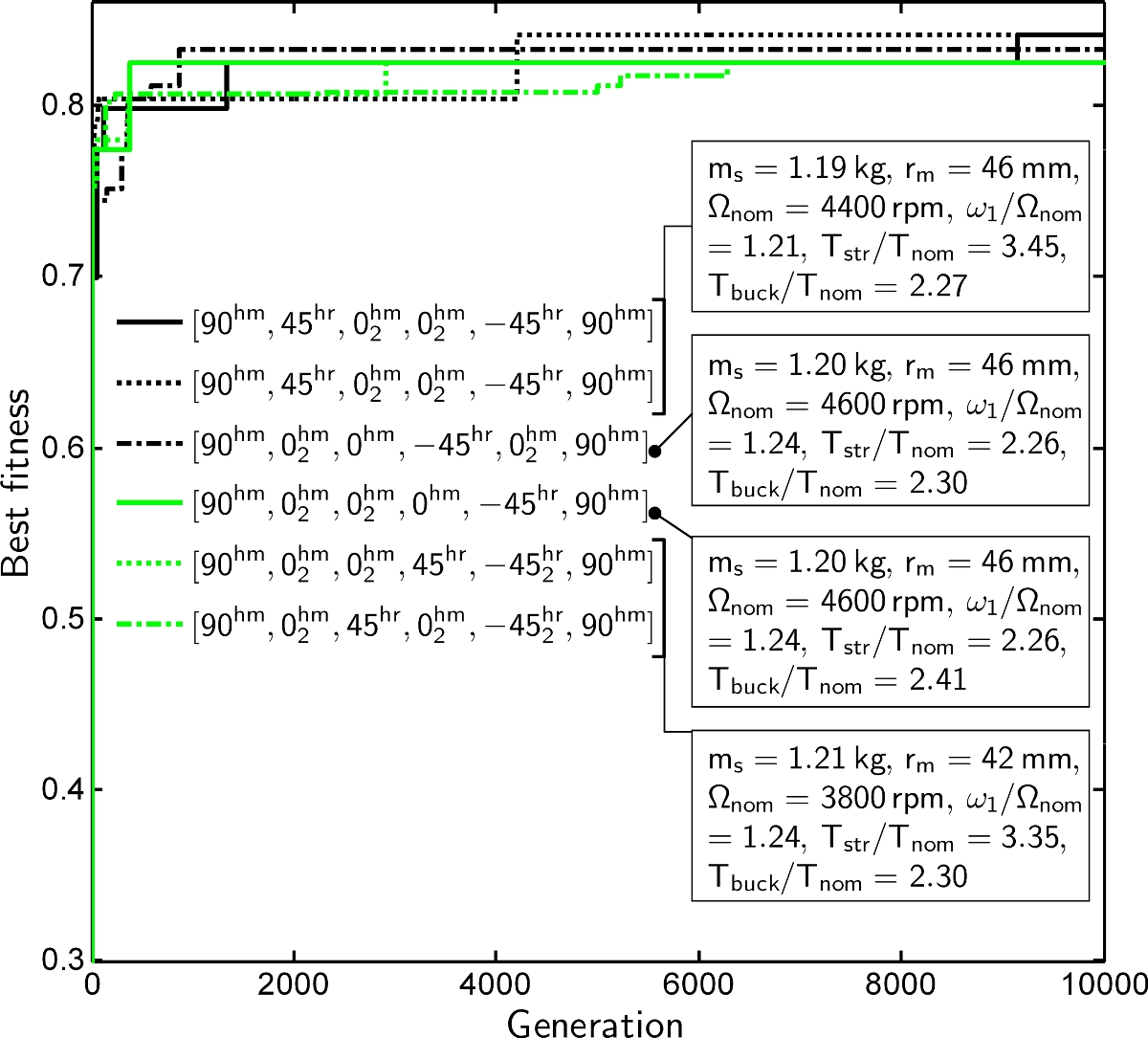}
\par\end{centering}

}
\par\end{centering}

\centering{}\caption{{\footnotesize Evolution of the best individual fitness of several
shaft populations in the case of various materials, subcritical conditions
and three tubes forming the \citeauthor{Zinberg70} tail rotor driveline
(there are 300 individuals in each evolution with $\mathrm{bit}_{n}=1$,
$\mathrm{bit}_{r_{\mathrm{m}}}=3$, $\mathrm{bit}_{\Omega_{\mathrm{nom}}}=3$,
$q=6$, $\Omega_{\mathrm{nom}}\in[3800,\,5200]$\,rev\,/\,min,
$t_{\mathrm{\srm\, min}}=1$\,mm and $l=$2.470\,m) \label{fig:souscritique}}}
\end{figure*}

{\small Subcritical shaft optimisation was first studied with the
GA. The case of two single-carbon fibre/epoxy composites (BE and HM)
and one hybrid case (HM/HS) was examined (see Table~\ref{tab:carbon_properties}
for the properties of the materials). The optimisation procedure was
computed six times with each composites to check that convergence
of the GA was reached. The results of the optimisation are shown in
Fig.~\ref{fig:souscritique} and the properties of the fittest individuals
are summarise in Table~\ref{tab:compare_shaft-sub} in each case.
Figs.~\ref{fig:sous-BE}-\ref{fig:sous-HM-HR} show the fitness function
of the highest ranking individual during the evolution (number of
generations) in six different populations. The legend gives the properties
of the fittest individual in each population at the last generation.
The results obtained show that the five conventional shafts can be
replaced by three subcritical boron or carbon reinforced epoxy shafts.
The algorithm did not find any subcritical solutions with only two
shafts. We note that the operating speed was much higher than the
first natural torsional frequency in all the solutions. }{\small \par}

{\small In the BE case (Fig.~\ref{fig:sous-BE}), GA yielded three
different solutions with the same fitness after 2000 generations.
All the solutions gave the same radius, the same operating speed and
the same plies (three 90\textdegree{} plies, four 0\textdegree{} plies,
one -45\textdegree{} ply and one 45\textdegree{} ply) but various
stacking sequence orders. They also gave the same critical speeds
and the same strength. The independence of critical speed computations
from the stacking sequence order is due to EMBT. As far as the strength
is concerned, this independence results from in-plane shear loading
and the assumption that uncoupled tension-bending is involved ($\mathbf{B}=\mathbf{0}$).
On the other hand, torsional buckling depends on the stacking sequence
order, as shown in Table~\ref{tab:buckling2}. In particular, the
circumferential flexural stiffness of the laminate was found to be
highly significant. This explains the position of the 90\textdegree{}
plies, which are located in the inner and outer parts of the tube.
The solution with the greatest buckling torque was selected as the
best individual. The stacking sequence obtained ({[}90\textdegree{}$_{2}$,
0\textdegree{}$_{4}$, 45\textdegree{}, -45\textdegree{}, 90\textdegree{}{]})
was very similar to that of \citeauthor{Zinberg70}'s laminate ({[}90\textdegree{},
45\textdegree{}, -45\textdegree{}, 0\textdegree{}$_{6}$, 90\textdegree{}{]}),
only two 0\textdegree{} plies were replaced here by a 90\textdegree{}
ply. The decrease in the shaft thickness and shaft radius explain
the slight increase in weight saving from 29\% to 30\% obtained in
comparison with the conventional aluminium shaft (see Table~\ref{tab:compare_shaft-sub}).
The computing time required for one evolution was approximately equal
to one hour using MATLAB \citep{Matlab90} on a Xeon E5540.}{\small \par}

{\small }{\small \par}

{\small The second material tested was HM car\-bon/e\-po\-xy (Fig.~\ref{fig:sous-HM}).
Convergence was reached after 200 generations, but five different
solutions were obtained with the same fitness. Only the order between
0\textdegree{}, 45\textdegree{} and -45\textdegree{} plies and the
operating speed were different. The optimum shaft stacking sequence
maximising the strength and buckling margins was {[}90\textdegree{},
0\textdegree{}$_{3}$, 45\textdegree{}, -45\textdegree{}$_{2}$, 45\textdegree{}{]}.
This gives the minimum thickness authorized (1\,mm). Due to the high
level of stiffness in comparison with BE, only one 90\textdegree{}
ply was necessary to prevent buckling and three 0\textdegree{} plies
were required to avoid reaching the first critical speed. The number
of $\pm$45\textdegree{} plies increased two-fold due to the low strength
of HM car\-bon/e\-po\-xy. The weight saving increased considerably
in comparison with the previous example, reaching 41\% due to several
combined effects: the decrease in the density, the mean tube radius,
the thickness and the weight of the supports (due to the increase
in the operating speed, see Eq.~(\ref{eq:masse_paliers})).}{\small \par}

{\small The optimisation of the HM car\-bon/e\-po\-xy was then
carried out with $\mathrm{bit}_{\alpha}=3$ i.e. $\alpha\in$\{-67.5\textdegree{},
-45\textdegree{}, -22.5\textdegree{}, 0\textdegree{}, 22.5\textdegree{},
45\textdegree{}, 67.5\textdegree{}, 90\textdegree{}\} (Fig.~\ref{fig:sous-HM-1}).
The chromosome length increased from 24 to 30. This considerably increased
the search-space, and hence the number of generations required to
obtain convergence and the computing time (approximately 3h/evolution).
Convergence was obtained after approximately 6000 generations. Four
different optimum individuals were obtained with the same fitness.
All of them contained (+ and -)22.5\textdegree{} plies, and most of
them contained (+ and/or -)67.5\textdegree{} plies. The optimum shaft
selected from four solutions was {[}90\textdegree{},-$23\text{\textdegree}_{2}$,23\textdegree{},-23\textdegree{},$23\text{\textdegree}_{2}$,-68\textdegree{}{]}.
This shaft did not contain 0\textdegree{} plies. The weight saving
increased slightly in comparison with the previous case, mainly due
to the decrease in the mean tube radius.}{\small \par}

{\small The last case tested was that of the hybrid HM/HS car\-bon/e\-po\-xy
(Fig.~\ref{fig:sous-HM-HR}). Convergence was again obtained after
approximately 6000 generations, despite the fact that only three evolutions
yielded the optimum individual. The optimum stacking sequence obtained
was {[}$90{^\circ}^{\mathrm{HM}}$, $45{^\circ}^{\mathrm{HR}}$, $0{^\circ}_{4}^{\mathrm{HM}}$,
-$45{^\circ}^{\mathrm{HR}}$, $90{^\circ}^{\mathrm{HM}}${]}. This
result requires some simple comments. The 90\textdegree{} and 0\textdegree{}
plies consisted of HM car\-bon/e\-po\-xy because these plies determine
the stiffness problems (the dynamics and buckling). The $\pm$45\textdegree{}
plies consisted of HS fibres because these plies determine the strength
problem. The weight reduction obtained in comparison with the HM case
was lower than expected. In fact, the decrease in the weight of the
shaft was practically balanced by the increase in the weight of the
supports.}{\small \par}

{\small }{\small \par}

{\small The optimisation procedure was also carried out in the case
of HS car\-bon/e\-po\-xy material (results not presented here).
In the configuration studied here, the HS car\-bon/e\-po\-xy material
gave a fitness score in between that obtained with BE and HM composite
materials, due to its low density. }{\small \par}

{\small In addition, it is worth noting that the number of generations
required to converge on the global optimum depended on the size of
the search-space, as well as on the basin of attraction of the local
and global optima. For example in the case of two materials with the
same chromosome length, the number of generations required to reach
convergence increased from 250 to 2000 (Figs.~\ref{fig:sous-BE}-\ref{fig:sous-HM}).}{\small \par}

{\small }{\small \par}

\begin{table*}[t]
\caption{{\footnotesize Optimised composite tail rotor driveline under supercritical
conditions in comparison with the conventional aluminium driveline
\label{tab:compare_shaft-super}}}

\begin{centering}
{\footnotesize }%
\begin{tabular}{llcccccc}
\hline 
 &  &  & {\footnotesize Conv.} & {\footnotesize Lim} & \multicolumn{3}{c}{}\tabularnewline
\cline{4-8} 
{\footnotesize Material} &  &  & {\footnotesize aluminium} & {\footnotesize CE$_{\mathrm{L.}}$\citep{Lim86}} & {\footnotesize HM} & {\footnotesize HM/HS} & {\footnotesize HM/HS}\tabularnewline
\hline 
{\footnotesize Number of tubes} &  & {\footnotesize -} & {\footnotesize 5} & {\footnotesize 1} & {\footnotesize 2} & {\footnotesize 2} & {\footnotesize 1 }\tabularnewline
\hline 
{\footnotesize String length} &  & {\footnotesize bit} & {\footnotesize -} & {\footnotesize -} & {\footnotesize 27} & {\footnotesize 33} & {\footnotesize 34}\tabularnewline
\hline 
{\footnotesize Stacking sequence} &  &  &  & {\footnotesize {[}$0_{64\%}$,} & {\footnotesize {[}$90$,$45$,} & {\footnotesize {[}$90^{\mathrm{HM}}$,$0_{3}^{\mathrm{HM}}$,} & {\footnotesize {[}$90^{\mathrm{HR}}$,$0_{9}^{\mathrm{HM}}$,}\tabularnewline
{\footnotesize (from inner to } &  & {\footnotesize \textdegree{}} & {\footnotesize -} & {\footnotesize $-62_{4\%}$,} & {\footnotesize $0_{2}$,$-45_{2}$,} & {\footnotesize $-45^{\mathrm{HR}}$,$0_{2}^{\mathrm{HM}}$,} & {\footnotesize $-45^{\mathrm{HR}}${]}}\tabularnewline
{\footnotesize outer radius)} &  &  &  & {\footnotesize $90_{32\%}${]}$_{s}$} & {\footnotesize $0$,$45${]} } & {\footnotesize $90^{\mathrm{HM}}${]} } & \tabularnewline
\hline 
{\footnotesize Operating speed} & {\footnotesize $\Omega_{\mathrm{nom}}$} & {\footnotesize rev\,/\,min} & {\footnotesize 5\,540} & {\footnotesize 6000} & {\footnotesize 5\,400} & \textcolor{black}{\footnotesize 4\,800} & {\footnotesize 7\,000}\tabularnewline
{\footnotesize 1st critical speed} & {\footnotesize $\omega_{1}$} & {\footnotesize rev\,/\,min} & {\footnotesize 8\,887} & {\footnotesize 490} & {\footnotesize 2\,696} & \textcolor{black}{\footnotesize 2\,647} & {\footnotesize 1\,018}\tabularnewline
{\footnotesize 2nd critical speed} & {\footnotesize $\omega_{2}$} & {\footnotesize rev\,/\,min} & {\footnotesize -} & {\footnotesize 1\,913}\textcolor{black}{\footnotesize $^{a}$} & {\footnotesize 10\,784} & \textcolor{black}{\footnotesize 10\,589} & {\footnotesize 4\,072}\tabularnewline
{\footnotesize 3nd critical speed} & {\footnotesize $\omega_{3}$} & {\footnotesize rev\,/\,min} & {\footnotesize -} & {\footnotesize 4\,303}\textcolor{black}{\footnotesize $^{a}$} & {\footnotesize 24\,264} & \textcolor{black}{\footnotesize 23\,824} & {\footnotesize 9\,161}\tabularnewline
{\footnotesize 4th critical speed} & {\footnotesize $\omega_{4}$} & {\footnotesize rev\,/\,min} & {\footnotesize -} & {\footnotesize 7\,650}\textcolor{black}{\footnotesize $^{a}$} & {\footnotesize 43\,136} & \textcolor{black}{\footnotesize 42\,355} & {\footnotesize 16\,287}\tabularnewline
{\footnotesize 1st torsion mode} & {\footnotesize $\varpi_{1}$} & {\footnotesize rev\,/\,min} & {\footnotesize 2\,058}\textcolor{black}{\footnotesize $^{a}$} & {\footnotesize 389}\textcolor{black}{\footnotesize $^{a}$} & {\footnotesize 1\,322} & \textcolor{black}{\footnotesize 409} & {\footnotesize 483}\tabularnewline
{\footnotesize 2nd torsion mode} & {\footnotesize $\varpi_{2}$} & {\footnotesize rev\,/\,min} & {\footnotesize 65\,370}\textcolor{black}{\footnotesize $^{a}$} & {\footnotesize 8\,846}\textcolor{black}{\footnotesize $^{a}$} & {\footnotesize 43\,326} & \textcolor{black}{\footnotesize 18\,112} & {\footnotesize 8\,300}\tabularnewline
{\footnotesize Threshold speed} & {\footnotesize $\omega_{\mathrm{th}}$} & {\footnotesize rev\,/\,min} & {\footnotesize -} & {\footnotesize -}\textcolor{black}{\footnotesize $^{b}$} & {\footnotesize 23\,658} & \textcolor{black}{\footnotesize 20\,356} & {\footnotesize 13\,638}\tabularnewline
\hline 
{\footnotesize Nominal torque} & {\footnotesize $T_{\mathrm{nom}}$} & {\footnotesize N\,m} & {\footnotesize 771} & {\footnotesize 712} & {\footnotesize 791} & \textcolor{black}{\footnotesize 891} & {\footnotesize 610}\tabularnewline
{\footnotesize Strength torque} & {\footnotesize $T_{\mathrm{str}}$} & {\footnotesize N\,m} & \textcolor{black}{\footnotesize 4\,925$^{a}$} & {\footnotesize 1\,492}\textcolor{black}{\footnotesize $^{a}$} & {\footnotesize 2\,439} & \textcolor{black}{\footnotesize 2\,096} & {\footnotesize 4\,352}\tabularnewline
{\footnotesize Buckling torque} & {\footnotesize $T_{\mathrm{buck}}$} & {\footnotesize N\,m} & \textcolor{black}{\footnotesize 3\,090$^{a}$} & {\footnotesize 1\,460}\textcolor{black}{\footnotesize $^{a}$} & {\footnotesize 1\,963} & \textcolor{black}{\footnotesize 2\,137} & {\footnotesize 1\,657}\tabularnewline
\hline 
{\footnotesize Tube length} & {\footnotesize $l$} & {\footnotesize m} & {\footnotesize 1.482} & {\footnotesize 7.41} & {\footnotesize 3.705} & \textcolor{black}{\footnotesize 3.705} & {\footnotesize 7.41}\tabularnewline
{\footnotesize Mean tube radius} & {\footnotesize $r_{\mathrm{m}}$} & {\footnotesize mm} & {\footnotesize 56.3} & {\footnotesize 47.7} & {\footnotesize 56.0} & \textcolor{black}{\footnotesize 50.0} & {\footnotesize 62.0}\tabularnewline
{\footnotesize Tube thickness} & {\footnotesize $t_{\mathrm{s}}$} & {\footnotesize mm} & {\footnotesize 1.65} & {\footnotesize 1.69} & {\footnotesize 1.0} & \textcolor{black}{\footnotesize 1.0} & {\footnotesize 1.375}\tabularnewline
\hline 
{\footnotesize Support stiffness} & {\footnotesize $k_{\mathrm{e}}$} & {\footnotesize kN~m$^{-1}$} & {\footnotesize -} & {\footnotesize -}\textcolor{black}{\footnotesize $^{b}$} & {\footnotesize 2\,864} & \textcolor{black}{\footnotesize 2\,864} & {\footnotesize 1\,437}\tabularnewline
\hline 
{\footnotesize Tubes weight} & {\footnotesize $N_{\srm}m_{\mathrm{s}}$} & {\footnotesize kg (\%)} & {\footnotesize 13.38} & {\footnotesize 6.08 (45)} & {\footnotesize 4.43 (33)} & \textcolor{black}{\footnotesize 3.60 (27)} & {\footnotesize 6.65 (50)}\tabularnewline
{\footnotesize Supports weight} & {\footnotesize $N_{\brm}m_{\mathrm{b}}$} & {\footnotesize kg} & {\footnotesize 15.42} & {\footnotesize 0} & {\footnotesize 3.80} & \textcolor{black}{\footnotesize 4.12} & {\footnotesize 0}\tabularnewline
{\footnotesize Weight penalty} &  & {\footnotesize kg} & {\footnotesize -} & {\footnotesize 1.5} & {\footnotesize 3.0} & \textcolor{black}{\footnotesize 3.0} & {\footnotesize 1.5}\tabularnewline
{\footnotesize Total weight} & {\footnotesize $m_{\mathrm{dv}}$} & {\footnotesize kg} & {\footnotesize 28.80} & {\footnotesize 7.58}\textcolor{black}{\footnotesize $^{a}$} & {\footnotesize 11.23} & \textcolor{black}{\footnotesize 10.72} & {\footnotesize 8.15 }\tabularnewline
{\footnotesize Weight saving} &  & {\footnotesize kg (\%)} & {\footnotesize -} & {\footnotesize 21.22 (74)} & {\footnotesize 17.6 (61)} & \textcolor{black}{\footnotesize 18.1 (63)} & {\footnotesize 20.7 (72)}\tabularnewline
\hline 
\end{tabular}
\par\end{centering}{\footnotesize \par}

\centering{}\textcolor{black}{\footnotesize $^{a}$~Value computed
with presented methods. $^{b}$~Not under consideration in the reference.}
\end{table*}

\begin{figure*}[!t]
\begin{centering}
\subfloat[{{\footnotesize HM ($\mathrm{bit}_{\mathrm{mat}}=0$, $r_{\mathrm{m}}\in$
{[}0.046,~0.06{]}\,mm, $\Omega_{\mathrm{nom}}\in${[}4800,~6200{]}\,rev\,/\,min)
\label{fig:sur-HM}}}]{\centering{}\includegraphics[height=7.2cm]{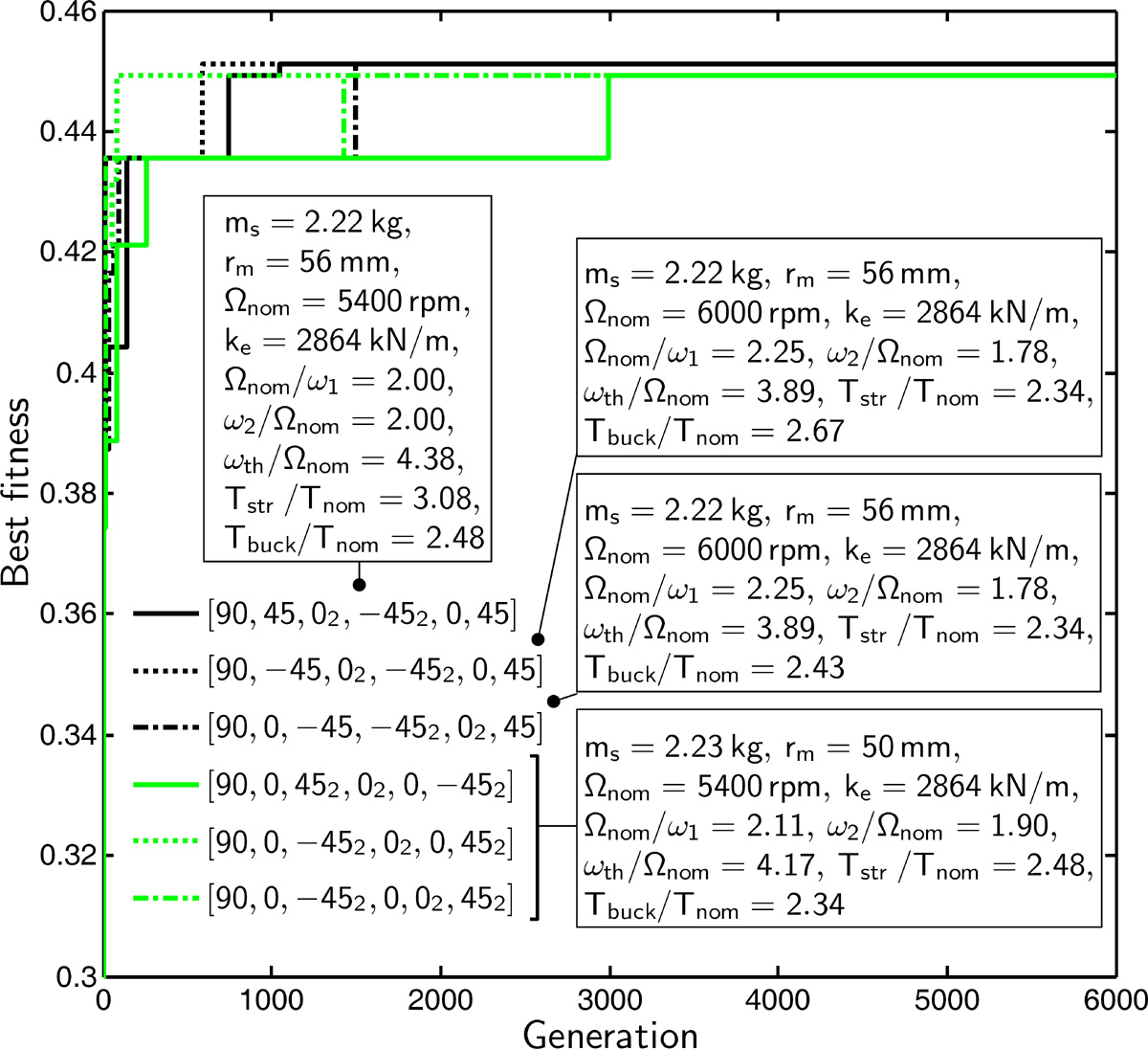}}\quad{}\subfloat[{{\footnotesize HM/HS ($\mathrm{bit}_{\mathrm{mat}}=1$, $r_{\mathrm{m}}\in$
{[}0.046,~0.06{]}\,mm, $\Omega_{\mathrm{nom}}\in${[}4800,~6200{]}\,rev\,/\,min)\label{fig:sur-HM-HR}}}]{\centering{}\includegraphics[height=7.2cm]{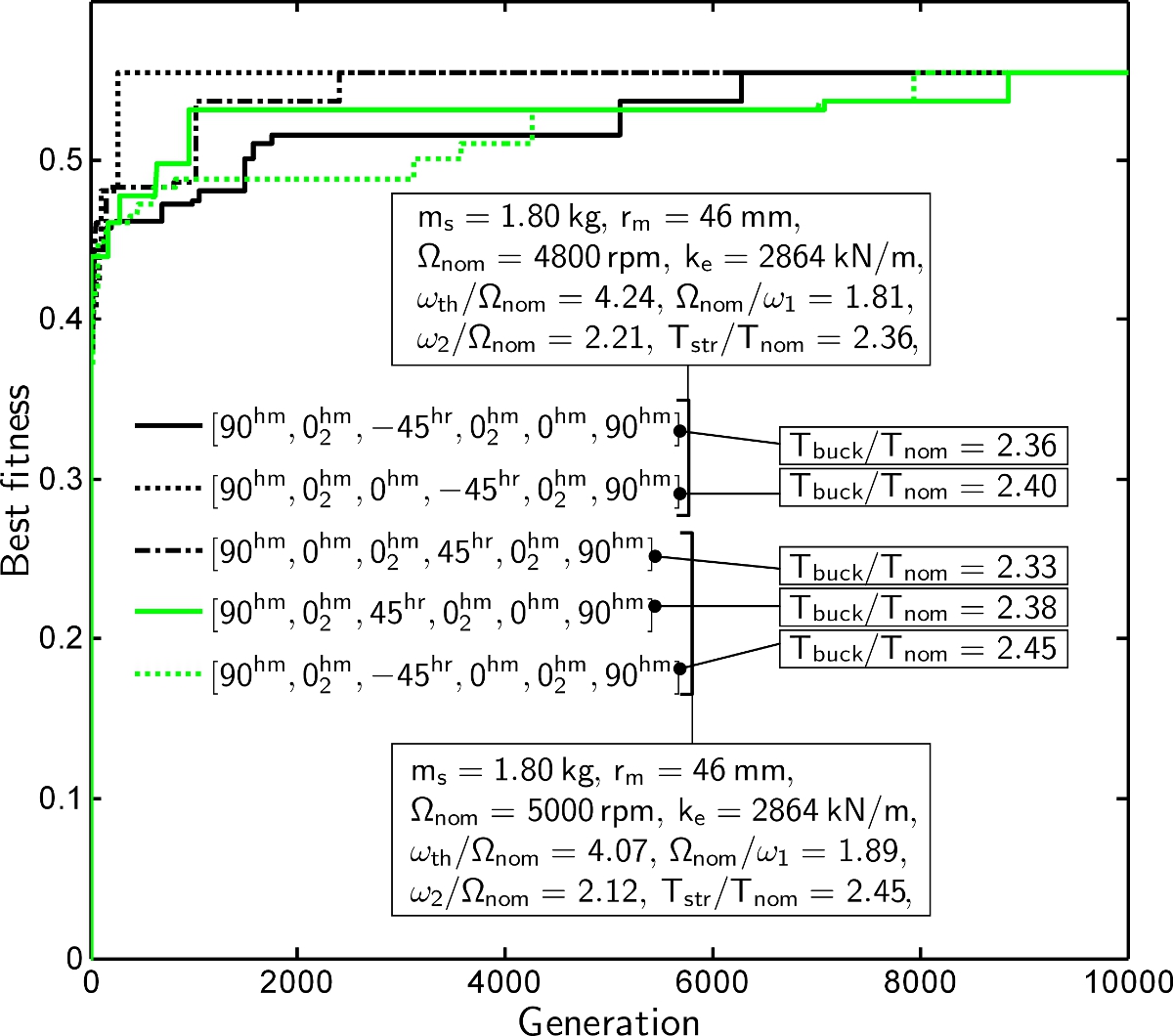}}
\par\end{centering}

\centering{}\caption{{\footnotesize Evolution of the best individual fitness of several
shaft populations in the case of various materials, supercritical
conditions and two tubes forming the \citeauthor{Zinberg70} tail
rotor driveline (there are 300 individuals in each evolution with
$\mathrm{bit}_{\alpha}=2$, $\mathrm{bit}_{n}=1$, $\mathrm{bit}_{r_{\mathrm{m}}}=3$,
$\mathrm{bit}_{k_{\erm}}=3$, $\mathrm{bit}_{\Omega_{\mathrm{nom}}}=3$,
$q=6$, $k_{\mathrm{e}}\in[10^{4},\,10^{7}]$\,N~m$^{-1}$, $t_{\mathrm{\srm\, min}}=1$\,mm,
 $\eta_{\mathrm{e}}=0.1$ and $l=$3.705\,m) \label{fig:surcritique-court}}}
\end{figure*}

\subsection{Supercritical shaft optimisation}

\begin{figure}[!t]
\begin{centering}
\includegraphics[height=7.2cm]{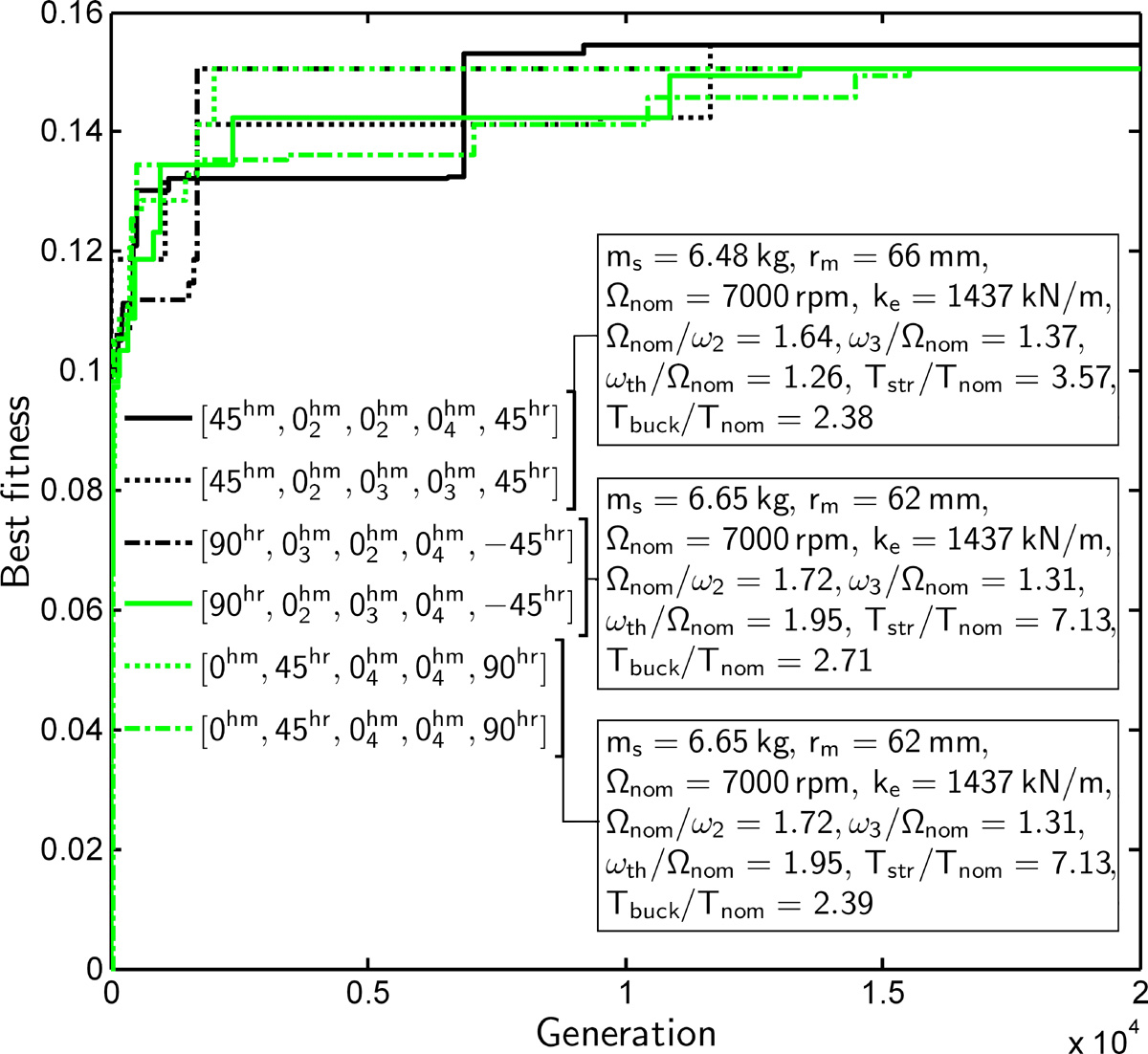}
\par\end{centering}

\centering{}\caption{{\footnotesize Evolution of the best individual fitness of several
shaft populations in the case of HM/HS composite material, supercritical
conditions and one tube forming the \citeauthor{Zinberg70} tail rotor
driveline (there are 600 individuals in each evolution with $\mathrm{bit}_{\alpha}=2$,
$\mathrm{bit}_{\mathrm{mat}}=1$, $\mathrm{bit}_{n}=2$, $\mathrm{bit}_{r_{\mathrm{m}}}=3$,
$\mathrm{bit}_{k_{\erm}}=3$, $\mathrm{bit}_{\Omega_{\mathrm{nom}}}=3$,
$q=5$, $r_{\mathrm{m}}\in[0.052,\,0.066]$\,mm, $\Omega_{\mathrm{nom}}\in[5600,\,7000]$\,rev\,/\,min,
$k_{\mathrm{e}}\in[10^{4},\,10^{7}]$\,N~m$^{-1}$, $t_{\mathrm{\srm\, min}}=1$\,mm,
 $\eta_{\mathrm{e}}=0.1$ and $l=$7.410\,m) \label{fig:sur-HM-HR-long}}}
\end{figure}

{\small The subcritical condition was then removed. The optimisation
was performed in the case of one single-carbon fibre/epoxy composite
(HM) and one hybrid composite (HM/HS). The results obtained here show
that the five conventional shafts can be replaced by either one or
two supercritical shafts (see Table~\ref{tab:compare_shaft-super}
and Figs.~\ref{fig:surcritique-court}-\ref{fig:sur-HM-HR-long}).
Contrary to the subcritical optimisation, it was necessary here to
take the first four critical speeds and the threshold speed into account.
The support stiffness was used as a supplementary optimisation variable
to maximise the dynamic stability margin. \citet{Lim86} suggested
optimising one shaft case with a car\-bon/e\-po\-xy composite denoted
here CE$_{\mathrm{L.}}$(Table~\ref{tab:carbon_properties}). The
latter authors used a generalised reduced gradient method involving
continuous variables such as the fraction and the orientation of the
laminate plies. Among the stacking sequences tested, {[}$0{^\circ}_{\alpha}$,
$\phi_{\beta p}$, $-\phi_{\beta n}$, $90{^\circ}_{\gamma}]_{s}$,
the optimum one was {[}$0{^\circ}_{64\%}$, $-62{^\circ}_{4\%}$,
$90{^\circ}_{32\%}]_{s}$ (Table~\ref{tab:compare_shaft-super}).
The operating speed was above the third critical speed and the dynamic
stability was not taken into account.}{\small \par}

{\small }{\small \par}

{\small In the two-tube case, the first material studied was HM car\-bon/e\-po\-xy
(Fig.~\ref{fig:sur-HM}). Convergence was reached with three populations
after approximately 6\,000 generations. The stacking sequence in
the optimum shaft (involving larger margins) was $[90{^\circ},$ $45{^\circ}$,
$0{^\circ}_{2}$,$-45{^\circ}_{2},$ $0{^\circ}$, $45{^\circ}${]}.
This sequence is the same as in the subcritical case, only the order
is slightly different. The shaft radius increased from 54 to 56\,mm.
It can be seen from Fig.~\ref{fig:sur-HM} that all the margins were
particularly large ($\geq2$). In particular, the operating speed
was in between the first and second critical speeds, and far above
the threshold speed. Much greater weight saving was obtained here
than with the conventional aluminium driveline (61\%) or the optimum
HM subcritical driveline, mainly due to the removal of one intermediate
support.}{\small \par}

{\small The second case tested was that of the hybrid HM/HS car\-bon/e\-po\-xy
(Fig.~\ref{fig:sur-HM-HR}). Five populations converged onto the
optimum fitness after 10\,000 generations. All the stacking sequences
consisted of five 0\textdegree{} plies made of HM car\-bon/e\-po\-xy,
two 90\textdegree{} plies made of HM carbon/epoxy, and one (+ or -)45\textdegree{}
ply made of HS car\-bon/e\-po\-xy. The fittest individual was {[}$90{^\circ}^{\mathrm{HM}}$,
$0{^\circ}_{3}^{\mathrm{HM}}$, -$45{^\circ}^{\mathrm{HR}}$, $0{^\circ}_{2}^{\mathrm{HM}}$,
$90{^\circ}^{\mathrm{HM}}${]}. This outcome is similar to that obtained
in the subcritical case, only one $45{^\circ}^{\mathrm{HR}}$ ply
was replaced by one $0{^\circ}^{\mathrm{HM}}$ ply. The weight saving
obtained was greater than in the HM fibre case, reaching 63\%.}{\small \par}

{\small }{\small \par}

{\small In the one-tube case, only hybrid HM/HS car\-bon/e\-po\-xy
was studied (Fig.~\ref{fig:sur-HM-HR-long}). Convergence was reached
with five populations after approximately 20\,000 generations. The
stacking sequence corresponding to the optimum individual was {[}$90{^\circ}^{\mathrm{HR}}$,
$0{^\circ}_{9}^{\mathrm{HM}}$, -$45{^\circ}^{\mathrm{HR}}${]}. It
is worth noting that the number of 0\textdegree{} plies increased
considerably in comparison with the two-tube case from 5 to 9, mainly
due to the fact that the dynamic constraints had to be achieved. The
operating speed here was in between the second and third critical
speeds. The AG selected HS fibres for the 90\textdegree{} ply here
instead of HM fibres, which is unusual, possibly because the tube
thickness was larger than in the previous cases tested, which reduced
the buckling risk. The weight saving amounted here to 72\%, which
is almost equal to that obtained by \citet{Lim86}. However, this
solution is more efficient because the operating speed was above the
second critical speed and the dynamic stability is ensured.}{\small \par}

{\small The above optimisation procedure was carried out with $\mathrm{bit}_{\alpha}=3$.
Even after 40\,000 generations with 8 populations of 600 individuals,
AG did not come up with a better solution. In this case, the computing
time was much longer, amounting to approximately 18\,hours per population.}{\small \par}

\section{Conclusion}

{\small In this study, some assumptions and simplifications were adopted
in order to describe the supercritical motion, the failure strength
and the torsional buckling of a CFRP drive shaft sufficiently accurately.
The GA presented for optimising supercritical drive shafts was tested
on an example previously described in the literature. Analytical models
are useful means of obtaining quite reasonable computing times. This
example shows the value of CFRP shafts and in particular, that of
hybrid CFRP shafts. These solutions make it possible to greatly decrease
the number of shafts and the driveline weight under subcritical conditions
and even more under supercritical conditions. In most of the cases
studied, the following general rules emerged for defining the stacking
sequence of hybrid solutions with\-out requiring any optimisation algorithms: }{\small \par}
\begin{enumerate}
\item {\small $\pm$45\textdegree{} HS carbon/epoxy plies should be used
in order to maximise the torque resistance, in variable proportions
ranging between +45\textdegree{} and -45\textdegree{}, depending on
the maximum torque direction; }{\small \par}
\item {\small 0\textdegree{} HM carbon/epoxy plies should be used in order
to maximise the axial stiffness and minimise the axial damping involved
in bending oscillations; }{\small \par}
\item {\small 90\textdegree{} HM carbon/epoxy plies should be used far from
the shaft middle surface in order to maximise the torsional buckling
torque;}{\small \par}
\item {\small the laminate does not generally have to be symmetrical.}{\small \par}
\end{enumerate}
\appendix

\section{Torsional buckling equations \label{sec:Torsional-buckling-equations}}

{\small Equilibrium equations used to solve the torsional buckling
problem in the case of a circular cylinder with orthotropic properties:
\begin{gather}
\left(A_{11}+\frac{B_{11}}{r}\right)u''+\left(2A_{13}-\frac{T}{\pi r^{2}}\right)\dot{u}'\nonumber \\
+\left(A_{33}-\frac{B_{33}}{r}+\frac{D_{33}}{r^{2}}\right)\ddot{u}+\left(A_{13}+\frac{2B_{13}}{r}+\frac{D_{13}}{r^{2}}\right)v''\nonumber \\
+\left(A_{12}+\frac{B_{12}}{r}+A_{33}+\frac{B_{33}}{r}\right)\dot{v}'+A_{23}\ddot{v}-\left(\frac{B_{11}}{r}+\frac{D_{11}}{r^{2}}\right)w'''\nonumber \\
+A_{12}w'+\left(-\frac{B_{23}}{r}+\frac{D_{23}}{r^{2}}\right)\dddot{w}+\left(A_{23}-\frac{B_{23}}{r}+\frac{D_{23}}{r^{2}}\right)\dot{w}\nonumber \\
+\left(-\frac{B_{12}}{r}-\frac{2B_{33}}{r}+\frac{D_{33}}{r^{2}}\right)\ddot{w}'-\left(\frac{3B_{13}}{r}+\frac{D_{13}}{r^{2}}\right)\dot{w}''=0\label{eq:buckling1}
\end{gather}
\begin{gather}
\left(A_{13}+\frac{2B_{13}}{r}+\frac{D_{13}}{r^{2}}\right)u''+\left(A_{12}+\frac{B_{12}}{r}+A_{33}+\frac{B_{33}}{r}-\frac{D_{33}}{2r^{2}}\right)\dot{u}'\nonumber \\
+\left(A_{23}+\frac{3D_{23}}{2r^{2}}\right)\ddot{u}+\left(A_{33}+\frac{3B_{33}}{r}+\frac{5D_{33}}{2r^{2}}\right)v''\nonumber \\
+\left(2A_{23}+\frac{4B_{23}}{r}+\frac{2D_{23}}{r^{2}}-\frac{T}{\pi r^{2}}\right)\dot{v}'+\left(A_{22}+\frac{B_{22}}{r}\right)\ddot{v}+A_{22}\dot{w}\nonumber \\
-\left(\frac{B_{13}}{r}+\frac{2D_{13}}{r^{2}}\right)w'''+\left(A_{23}+\frac{B_{23}}{r}-\frac{T}{\pi r^{2}}\right)w'-\frac{B_{22}}{r}\dddot{w}\nonumber \\
+\left(-\frac{B_{12}}{r}-\frac{D_{12}}{r^{2}}-\frac{2B_{33}}{r}-\frac{3D_{33}}{r^{2}}\right)\dot{w}''+\left(-\frac{3B_{23}}{r}+\frac{D_{23}}{r^{2}}\right)\ddot{w}'=0
\end{gather}
\begin{gather}
\left(\frac{B_{11}}{r}+\frac{D_{11}}{r^{2}}\right)u'''+\left(3\frac{B_{13}}{r}+\frac{D_{13}}{r^{2}}\right)\dot{u}''\nonumber \\
+\left(2\frac{B_{33}}{r}-\frac{D_{33}}{r^{2}}+\frac{B_{12}}{r}\right)\ddot{u}'+\left(\frac{B_{23}}{r}-\frac{D_{23}}{r^{2}}\right)\dddot{u}-A_{12}u'\nonumber \\
+\left(-A_{23}+\frac{B_{23}}{r}-\frac{D_{23}}{r}\right)\dot{u}+\left(\frac{B_{13}}{r}+\frac{2D_{13}}{r^{2}}\right)v'''\nonumber \\
+\left(2\frac{B_{33}}{r}+\frac{3D_{33}}{r^{2}}+\frac{B_{12}}{r}+\frac{D_{12}}{r^{2}}\right)\dot{v}''+\left(3\frac{B_{23}}{r}+\frac{2D_{23}}{r^{2}}\right)\ddot{v}'\nonumber \\
+\frac{B_{22}}{r}\dddot{v}+\left(-A_{23}-\frac{B_{23}}{r}+\frac{T}{\pi r^{2}}\right)v'-A_{22}\dot{v}-\frac{D_{11}}{r^{2}}w''''\nonumber \\
-\frac{4D_{13}}{r^{2}}\dot{w}'''-\left(\frac{4D_{33}}{r^{2}}+2\frac{D_{12}}{r^{2}}\right)\ddot{w}''-\frac{4D_{23}}{r^{2}}\dddot{w}'\nonumber \\
-\frac{D_{22}}{r^{2}}\ddddot{w}+\frac{3B_{12}}{r}w''+\left(\frac{4B_{23}}{r}-\frac{2D_{23}}{r^{2}}-\frac{T}{\pi r^{2}}\right)\dot{w}'\nonumber \\
+\left(\frac{2B_{22}}{r}-\frac{2D_{22}}{r^{2}}\right)\ddot{w}+\left(-A_{22}+\frac{B_{22}}{r}-\frac{D_{22}}{r^{2}}\right)w=0\label{eq:buckling3}
\end{gather}
where $\_'=r\partial\_/\partial x$ and $\dot{\_}=\partial\_/\partial\varphi$.}{\small \par}

{\small  Elements of the stiffness matrix in the torsional buckling
problem in the case of a very long circular cylinder with orthotropic
properties: 
\begin{gather}
K(1,1)=-\left(A_{11}+\frac{B_{11}}{r}\right)\lambda^{2}-\left(2A_{13}-\frac{T}{\pi r^{2}}\right)h\lambda\nonumber \\
-\left(A_{33}-\frac{B_{33}}{r}+\frac{D_{33}}{r^{2}}\right)h^{2}\label{eq:Tbuck11}\\
K(1,2)=-\left(A_{13}+\frac{2B_{13}}{r}+\frac{D_{13}}{r^{2}}\right)\lambda^{2}\nonumber \\
-\left(A_{12}+\frac{B_{12}}{r}+A_{33}+\frac{B_{33}}{r}\right)h\lambda-A_{23}h^{2}\\
K(1,3)=-\left(\frac{B_{11}}{r}+\frac{D_{11}}{r^{2}}\right)\lambda^{3}+A_{12}\lambda\nonumber \\
+\left(-\frac{B_{23}}{r}+\frac{D_{23}}{r^{2}}\right)h^{3}+\left(A_{23}-\frac{B_{23}}{r}+\frac{D_{23}}{r^{2}}\right)h\nonumber \\
+\left(-\frac{B_{12}}{r}-\frac{2B_{33}}{r}+\frac{D_{33}}{r^{2}}\right)h^{2}\lambda-\left(\frac{3B_{13}}{r}+\frac{D_{13}}{r^{2}}\right)h\lambda^{2}\\
K(2,1)=-\left(A_{13}+\frac{2B_{13}}{r}+\frac{D_{13}}{r^{2}}\right)\lambda^{2}\nonumber \\
-\left(A_{12}+\frac{B_{12}}{r}+A_{33}+\frac{B_{33}}{r}-\frac{D_{33}}{2r^{2}}\right)h\lambda-\left(A_{23}+\frac{3D_{23}}{2r^{2}}\right)h^{2}\\
K(2,2)=-\left(A_{33}+\frac{3B_{33}}{r}+\frac{5D_{33}}{2r^{2}}\right)\lambda^{2}\nonumber \\
-\left(2A_{23}+\frac{4B_{23}}{r}+\frac{2D_{23}}{r^{2}}-\frac{T}{\pi r^{2}}\right)h\lambda-\left(A_{22}+\frac{B_{22}}{r}\right)h^{2}
\end{gather}
\begin{gather}
K(2,3)=-\left(\frac{B_{13}}{r}-\frac{2D_{13}}{r^{2}}\right)\lambda^{3}+\left(A_{23}+\frac{B_{23}}{r}-\frac{T}{\pi r^{2}}\right)\lambda\nonumber \\
-\frac{B_{22}}{r}h^{3}+A_{22}h+\left(-\frac{3B_{23}}{r}+\frac{D_{23}}{r^{2}}\right)h^{2}\lambda\nonumber \\
-\left(\frac{B_{12}}{r}+\frac{D_{12}}{r^{2}}+\frac{2B_{33}}{r}+\frac{3D_{33}}{r^{2}}\right)h\lambda^{2}\\
K(3,1)=-\left(\frac{B_{11}}{r}+\frac{D_{11}}{r^{2}}\right)\lambda^{3}-\left(3\frac{B_{13}}{r}+\frac{D_{13}}{r^{2}}\right)h\lambda^{2}\nonumber \\
-\left(2\frac{B_{33}}{r}-\frac{D_{33}}{r^{2}}+\frac{B_{12}}{r}h^{2}\right)\lambda\nonumber \\
-\left(\frac{B_{23}}{r}-\frac{D_{23}}{r^{2}}\right)h^{3}-A_{12}\lambda+\left(-A_{23}+\frac{B_{23}}{r}-\frac{D_{23}}{r}\right)h\\
K(3,2)=-\left(\frac{B_{13}}{r}+\frac{2D_{13}}{r^{2}}\right)\lambda^{3}\nonumber \\
-\left(2\frac{B_{33}}{r}+\frac{3D_{33}}{r^{2}}+\frac{B_{12}}{r}+\frac{D_{12}}{r^{2}}\right)h\lambda^{2}-\left(3\frac{B_{23}}{r}+\frac{2D_{23}}{r^{2}}\right)h^{2}\lambda\nonumber \\
-\frac{B_{22}}{r}h^{3}+\left(-A_{23}-\frac{B_{23}}{r}+\frac{T}{\pi r^{2}}\right)\lambda-A_{22}h\\
K(3,3)=-\frac{D_{11}}{r^{2}}\lambda^{4}-\frac{4D_{13}}{r^{2}}h\lambda^{3}-\left(\frac{4D_{33}}{r^{2}}+2\frac{D_{12}}{r^{2}}\right)h^{2}\lambda^{2}\nonumber \\
-\frac{4D_{23}}{r^{2}}h^{3}\lambda-\frac{D_{22}}{r^{2}}h^{4}-\frac{3B_{12}}{r}\lambda^{2}-\left(\frac{4B_{23}}{r}-\frac{2D_{23}}{r^{2}}-\frac{T}{\pi r^{2}}\right)h\lambda\nonumber \\
-\left(\frac{2B_{22}}{r}-\frac{2D_{22}}{r^{2}}\right)h^{2}+\left(-A_{22}+\frac{B_{22}}{r}-\frac{D_{22}}{r^{2}}\right)\label{eq:Tbuck33}
\end{gather}
where $\lambda=p\pi r/l$.}{\small \par}

{\small }{\small \par}

\bibliographystyle{elsarticle-num-names}
\bibliography{c:/MyLaTeX/Bib_comp,c:/MyLaTeX/Bib_struct,c:/MyLaTeX/Bib_rotor,c:/MyLaTeX/Bib_optim,C:/MyLaTeX/Bib_matlab}

\end{document}